\documentclass{egpubl}
\usepackage{eg2025}
 
\ConferencePaper        
\CGFccby

\usepackage[T1]{fontenc}
\usepackage{dfadobe}  

\usepackage{cite}  
\BibtexOrBiblatex
\electronicVersion
\PrintedOrElectronic
\usepackage[pdftex]{graphicx} 

\usepackage{egweblnk} 

\usepackage{amsfonts}
\usepackage{booktabs}
\usepackage{xspace}
\usepackage{amsmath}

\usepackage{ifthen}

\newboolean{showComment}
\setboolean{showComment}{true} 

\ifthenelse{\boolean{showComment}}{
    
\newcommand{\purvi}[1]{\textcolor{blue}{[purvi]: #1}}
\newcommand{\needcite}[1]{\textcolor{red}{[CITE: #1}}
\newcommand{\karen}[1]{\textcolor{red}{[Karen: #1]}}

\newcommand{\edit}[1]{\textcolor{black}{ #1}}
\newcommand{\mia}[1]{\textcolor{blue}{[mia]: #1}}
\newcommand{\vishnu}[1]{\textcolor{blue}{[vishnu]: #1}}
}{

\newcommand{\purvi}[1]{}
\newcommand{\KF}[1]{}
\newcommand{\needcite}[1]{}
\newcommand{\karen}[1]{}
\newcommand{\s}[1]{}
\newcommand{\edit}[1]{}
\newcommand{\mia}[1]{}
\newcommand{\vishnu}[1]{}
}

\newcommand{\blocking}{$\mathbf{X}$\xspace}

\newcommand{\detailed}{$\mathbf{Y}$\xspace}

\newcommand{\residual}{$\Delta \mathbf{X}$\xspace}

\newcommand{\functionwarp}{$\texttt{warp}(\mathbf{w}, \mathbf{X}$)\xspace}

\newcommand{\blockingpose}{$\mathbf{X}_{k+\Delta k}$\xspace}

\newcommand{\detailedpose}{$\mathbf{Y}_{k}$\xspace}

\title[Generative Motion Infilling from Imprecisely Timed Keyframes]%
      {Generative Motion Infilling from Imprecisely Timed Keyframes}

\author[P. Goel, H. Zhang, C. K. Liu, K. Fatahalian]
{\parbox{\textwidth}{\centering P. Goel$^{1}$\orcid{0000-0003-2618-092X} \
        H. Zhang$^{2}$\orcid{0009-0008-0293-337X} \
        C. K. Liu$^1$\orcid{0000-0001-5926-0905} \
        K. Fatahalian$^1$\orcid{0000-0001-8754-0429} 
        }
        \\
{\parbox{\textwidth}{\centering $^1$Stanford University
         $^2$NVIDIA
       }
}
}

\begin{document}


\teaser{
\centering
\includegraphics[trim={20 0 75 100},clip,width=\linewidth]{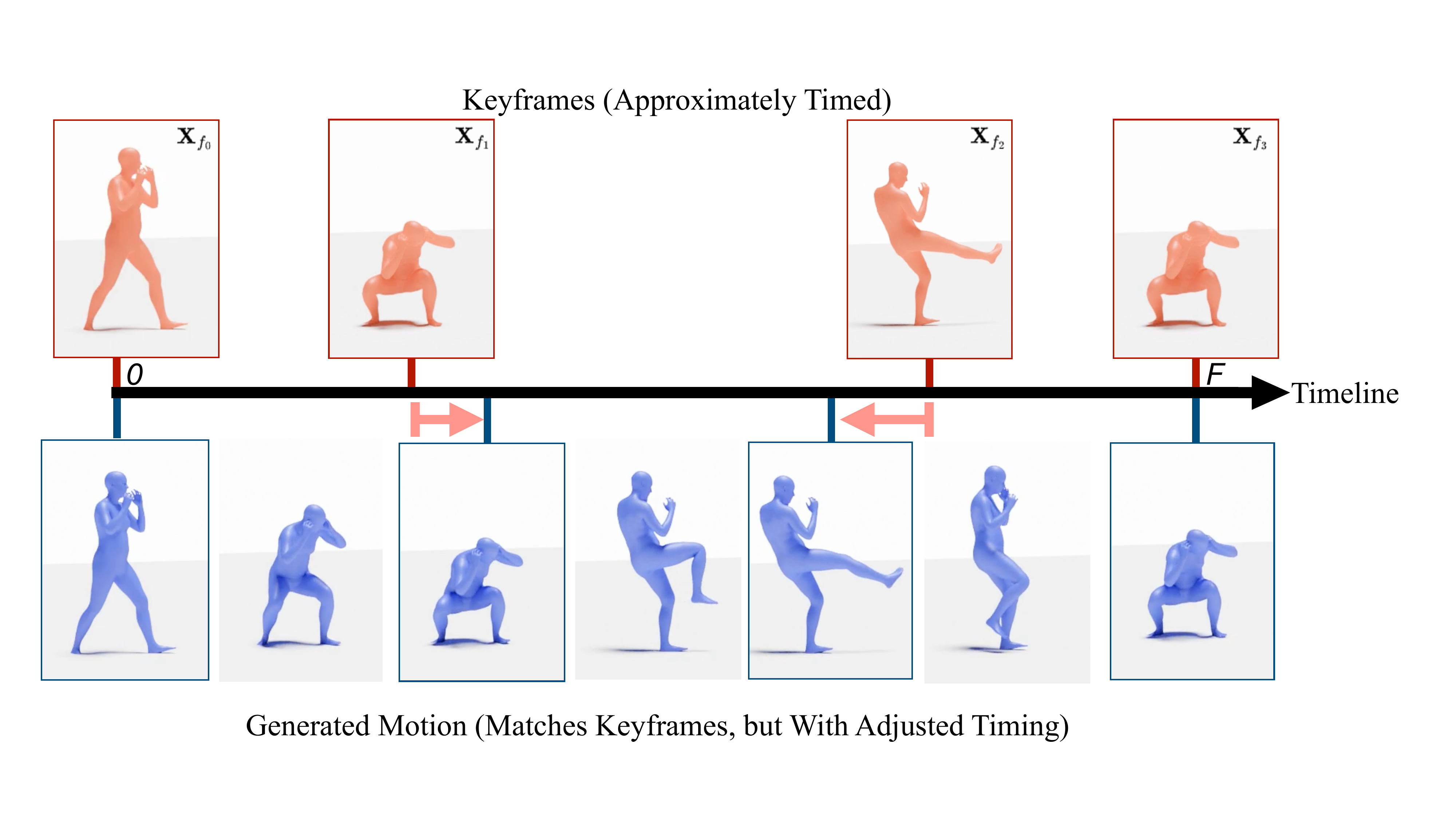}

\caption{ Our method performs motion inbetweening of keyframes (top, red) that \textit{may} be imprecisely timed. Here, the user creates four keyframes corresponding to the iconic moments in a desired martial arts sequence--``the character ducks, then kicks, then ducks.'' The keyframes are approximately timed, meaning they may not positioned on the timeline at the precise moments needed for the desired movement (e.g., because producing correctly timed keyframes is a tedious process requiring significant skill). Our method generates a high-fidelity motion sequence (bottom, blue), retiming the input keyframes (pink arrows) as necessary to ensure plausible timing while still adhering to the poses themselves (bottom, outlined). For example, the first ducking pose is pushed a few frames forward, which gives the character enough time to fully reach the pose. The final motion also contains spatial details between the keyframes, like a snappy kick and weight shifts.  }
\label{fig:banner}
}

\maketitle

\begin{abstract}
Keyframes are a standard representation for kinematic motion specification. Recent learned motion-inbetweening methods use keyframes as a way to control generative motion models, and are trained to generate life-like motion that matches the exact poses and timings of input keyframes. However, the quality of generated motion may degrade if the timing of these constraints is not perfectly consistent with the desired motion. Unfortunately, correctly specifying keyframe timings is a tedious and challenging task in practice.  Our goal is to create a system that synthesizes high-quality motion from keyframes, even if keyframes are imprecisely timed. We present a method that allows constraints to be retimed as part of the generation process. Specifically, we introduce a novel model architecture that explicitly outputs a time-warping function to correct mistimed keyframes, and spatial residuals that add pose details.  We demonstrate how our method can automatically turn approximately timed keyframe constraints into diverse, realistic motions with plausible timing and detailed submovements.

\end{abstract}

\maketitle

\section{Introduction}
\label{sec:intro}

Recent advancements in generative motion-inbetweening have demonstrated promising abilities for generating motion from keyframe constraints. These methods integrate keyframes as a control signal in the motion generation process, and are capable of generating natural motion that adheres to the given constraints.  

The problem is that while animators may be able to precisely specify the spatial component of keyframe constraints, e.g., how the joints are articulated, the process of correctly ``timing the animation'' (precisely positioning keyframe constraints on the timeline so that interpolation yields a desired result) has been found to  demand comparatively more animator skill. Terra et al. ~\cite{performancetiming} observe through interviews that timing an animation can be one of the most difficult parts of the process for novice users--considerably more difficult than posing. Experienced animators have observed that producing a detailed motion that arrives at a target pose even a few frames too soon or too late can significantly affect the meaning of the final motion ~\cite{animatorssurvivalkit, timingforanimation, principlesofanimation3d}. In other words, if the timing is ``wrong'', the interpolation of the keyframes may not match the desired result.

Existing motion inbetweening solutions are trained to match input keyposes at exactly the frame provided. In practice, this hard timing constraint does not pose much of a problem to learned motion-inbetweening models if there are only two or three keyframe constraints, since the model has significant flexibility to construct a motion in between the keyframes that still looks natural. But given that the model is very sparsely constrained, the generated motion--despite appearing natural and adhering to constraints--may  not reflect the final detailed animation that the animator envisioned. If the animator seeks more control and provides more keyframes, the model has less flexibility to compensate for mistimed keyframe inputs. In fact, there may be no such natural motion that meets the hard keyframe constraints. The result is generated output that features unrealistic dynamics (the character moves from one keyframe to another too fast) or even fails to hit keyframes (the character does not have enough time to reach the next keyframe). For this reason, \edit{in order to enable a flexible and user-friendly animation workflow}, we believe that a practical learned inbetweening system \edit{that supports motion synthesis and motion editing} must have the ability to adjust the timing of input keyframes.

In this paper, we design a motion-inbetweening model that synthesizes high quality motion, but does so in the context of keyframes that may only be approximately timed. Specifically, we make the following contributions.

\begin{itemize}
    \item A diffusion model architecture designed to transform imprecisely timed keyposes into high-fidelity, detailed motion. The model uses a dual-head approach, accounting for temporal imprecision in keyframe constraints by predicting both a global time warp of the input (to adjust timing) and local pose residuals (to add spatial motion detail).  
    \item A dataset generation scheme that creates plausible imprecisely-timed keyframes from detailed motion clips. These corresponding pairs of keyframe constraints and detailed motion serve as training data for learning how to perform motion infilling in the context of approximately-timed poses.
\end{itemize}

We demonstrate the system's ability to generate high-quality motion output from approximately timed keyframe constraints in both motion synthesis and motion editing tasks.

\section{Related Work}

\subsection{Motion In-betweening}
Closely related to our work is motion in-betweening, which generates a full motion sequence from a set of keyframe constraints. \edit{Machine learning methods have demonstrated excellent performance in generating high-quality motion from even very sparse keyframe constraints, leveraging, e.g., RNNs~\cite{harveypal2, 2018-MIG-autoComplete, harvey2020robust}, GANs~\cite{Ahn2017Text2ActionGA, Ghosh2021SynthesisOC}, Transformer-based architectures~\cite{twostagetransformers,faceinbetween,skelbetween,mo2023continuousintermediatetokenlearning}, and auto-encoders~\cite{learnedmotionmatching,Kaufmann_2020,oreshkin2022motion}. Importantly, the in-betweening task \textit{by definition} treats keyframe timings as hard constraints, and assumes the input timing is precise.} Recently some motion diffusion models have been proposed for the in-betweening task, accepting keypose constraints through, e.g., observation masks~\cite{goelsiggraph, wei2023enhanced, cohan2024flexible}, guidance~\cite{karunratanakul2023gmd, xie2024omnicontrol}, or inference-time imputation~\cite{tseng2022edge, tevet2023human}. These methods, too, assume input constraints have correct timing, and do not meet our goal to support inbetweening in the context where keyframe constraints may be imprecisely timed.

\subsection{Loose constraints in diffusion.}  
There is significant interest in developing improved ways to add interpretable control to generative models~\cite{maneesh:blackboxes:2023}. Specifically, we are inspired by recent work that develops mechanisms for artists to block out scene composition with coarse primitives~\cite{bhat2023loosecontrol} or convey the gist of a scene by drawing a simple sketch that is interpreted loosely by the generative model~\cite{sarukkai2024block}. Our goal of producing a method for generative motion infilling under imprecise timing constraints corresponds with the common animator observations about the difficulty and tediousness of providing perfectly timed keyframe constraints. 

Motion diffusion methods can generate high quality motion from conditions like text~\cite{zhang2022motiondiffuse, tevet2023human, dabral2023mofusion,chen2023executing,flexinbetween,shafir2023human, kapon2024mas, petrovich24stmc}, dense trajectory constraints~\cite{ karunratanakul2023dno, karunratanakul2023guided, rempeluo2023tracepace}, and music~\cite{tseng2022edge}. Many such models can be extended to allow ``loose'' interpretation of joint-level constraints, to some extent. 

For example, relevant to spatial constraints like keyframes, inference-time imputation techniques~\cite{tevet2023human,shafir2023human,goelsiggraph} involve replacing the output of some number of diffusion inference steps with a noisy version of the input constraint, e.g., inpainting desired keyframes for until some diffusion step \textit{C}. We argue that these techniques are designed to control how strongly the generated motion corresponds with the \textit{entire} condition signal. In the context of \textit{loose timing constraints}, only the \textit{timing} of the keyframe constraints should be considered a loose constraint. The poses themselves--which, in many practical use cases, are often meticulously crafted by artists--should stay as unchanged as possible. 

\subsection{Automatic Motion Retiming}

An almost universeral operator for retiming motion in traditional animation tools is the time warp operation, first proposed as a spline-based mapping by~\cite{Witkin1995MotionW}. Manually specifying time warp splines can be a tedious process and require significant domain expertise. Follow-up work has explored more automatic ways to synthesize timewarps (and other parameterizations for motion retiming), such as  acting \cite{performancetiming}, optimization \cite{oldtimewarp, physicsbasedretiming,relaxingtiming,familyofmotions}, and demonstration through a reference motion \cite{Metaxas2007GuidedTW}. Similar to many of these works, we parameterize keyframe retiming via a time warp; unlike these works, we seek to learn the space of plausible timings from a motion database, rather than rely on, e.g., hand-crafted optimization objectives or a single reference motion.

\section{Problem Description}
\label{prelim}

Our goal is to generalize the task of motion in-betweening to allow for more flexible timing control, where keyframe constraints need only be approximately timed. Traditionally, motion-inbetweening methods abide by the following problem statement. Given a set of \textit{N} keyframe constraints $K = \{ \mathbf{x}_{f_0}, \mathbf{x}_{f_1}, ... \mathbf{x}_{f_N}\}$, where the \textit{n}-th keyframe $\mathbf{x}_{f_n}$ specifies a keypose $\mathbf{x}_n$ located at frame index $f_n$, generate a motion $\mathbf{Y} = \{\mathbf{y}_i \}^F_{i=0}$ (where \textit{F} is the total number of frames) that adheres as closely as possible to the keyframe constraints. In other words,

\begin{equation}
\mathbf{x}_{f_n} = \mathbf{y}_{f_n}, \forall n \in {0, 1, ..., N}
\end{equation}

At the same time, $\mathbf{Y}$ should maintain coherence between the input keyframes and the rest of the motion, while containing the detail of human motion. As we have discussed, these goals can conflict in complex ways when the keyframe timing is imprecise.

We propose a more flexible approach to this task. The generated motion, $\mathbf{Y}$, should be a realistic and plausible human motion that remains coherent with the input keyframes. While the animation is expected to reach the key poses at some point in time, it does not necessarily need to match the exact timings specified in $\mathbf{X}$, i.e.,

\begin{equation}
\mathbf{x}_{f_n} = \mathbf{y}_{m}, \forall n \in {0, 1, ..., N} \ \text{and} \ m \in [f_n-P, f_n + P )
\end{equation}

where \textit{P} is a small integer.

When placed on the timeline,  \textit{K} yields a discontinuous motion that we refer to as the observation signal $\mathbf{X}$. $\mathbf{X}$ may constrained sparsely, i.e., just a few keyframes, or densely, i.e., $\mathbf{X}$ may contain high-fidelity submotions. $\mathbf{X}$ is undefined at unconstrained frames. The system converts observation signal $\mathbf{X}$ to a detailed, high-fidelity $\mathbf{Y}$ that adheres to the coarse structure of the keyframe constraints (e.g., it contains the same poses as the original keyframes, at similar points in time, but the timing need not be an exact match), while exhibiting realistic motion.

\section{Method}
\label{sec:datagen}

Many prior works on learned motion-inbetweening follow a common pattern: first, generate a synthetic dataset consisting of an observation signal $\mathbf{X}$ with corresponding high-fidelity motion $\mathbf{Y}$. Typically, $\mathbf{X}$ contains some keyframes sampled from $\mathbf{Y}$. A model is then trained to generate the complete $\mathbf{Y}$ from $\mathbf{X}$.

We adopt this approach but introduce two key innovations. First, we propose a new data generation procedure such that the keyframes in $\mathbf{X}$ are deliberately mistimed, simulating real-world inconsistencies. Second, we introduce a novel model architecture that jointly predicts (a) an explicit global time-warping function to correct and plausibly retime the mistimed keyframe constraints, and (b) local pose residuals that add spatial detail. Together, these innovations allow the model to  generate high-quality, realistic motion even in the presence of imprecise, approximate keyframe timing.

\subsection{Dataset Generation}
\label{sec:datasetgen}

\begin{figure}
    \centering
    \includegraphics[width=\linewidth]{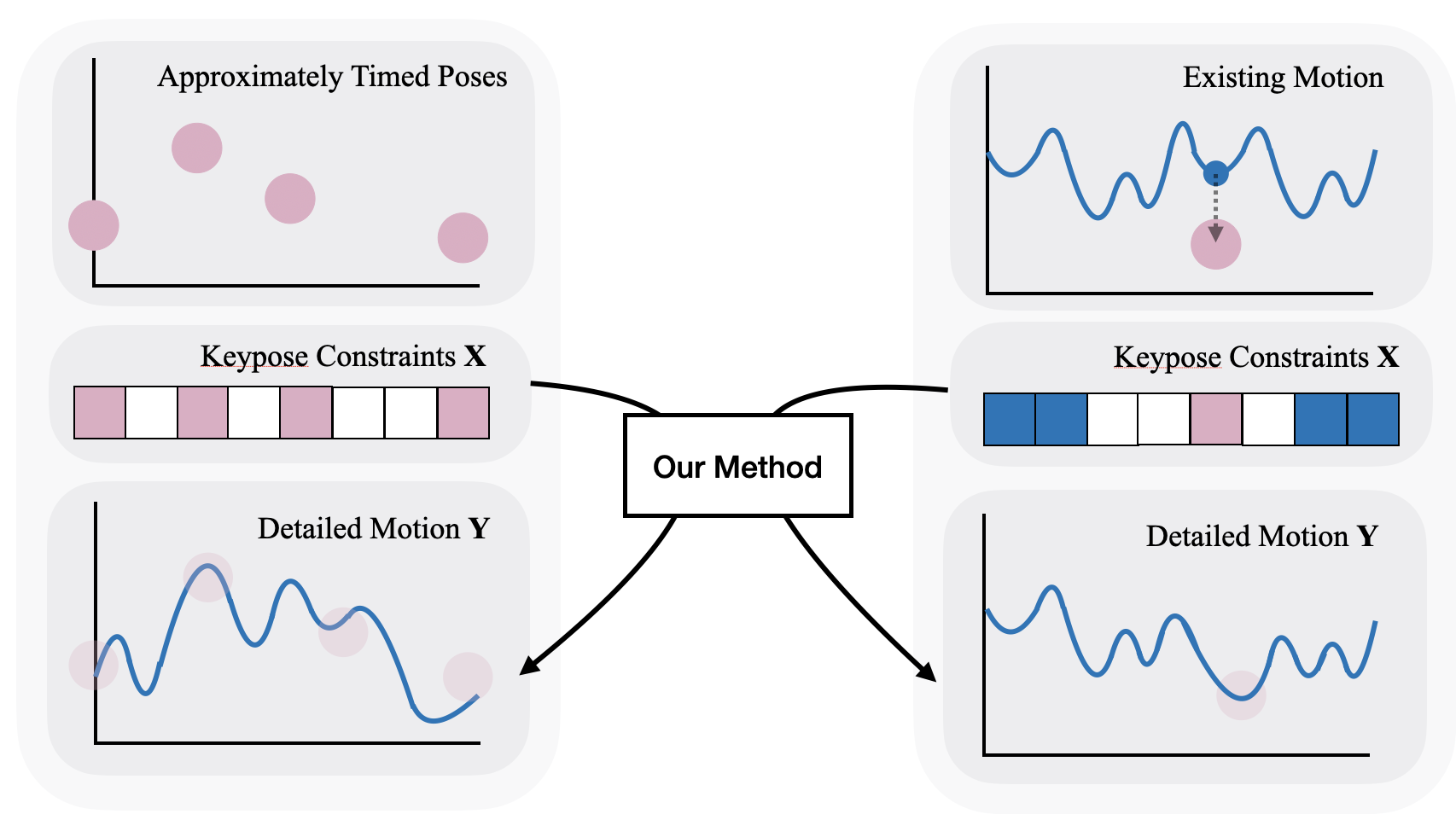}
    \caption{\textbf{System}: Our method for motion infilling with loose timing control accomodates motion synthesis (left), and motion editing (right). In the motion synthesis workflow, the animator provides a set of keyposes, and approximately when these events occur on the timeline (left, top). The union of constrained and unconstrained regions form the observation signal $\mathbf{X}$, which our method converts into detailed, high-fidelity motion \detailed (left, bottom). In the the motion editing workflow, the animator starts with an existing high-fidelity motion (right, top), and specifies an edit by providing a new keypose (right, top: pink dot). This can result in in observation signal $\mathbf{X}$ comprising context from the original motion and thew new keypose (right, middle). Then, as our method converts \blocking into \detailed by adding pose and timing detail (right, bottom).   } 
    \label{system}
\end{figure}

\begin{figure}
    \centering
    \includegraphics[width=\linewidth]{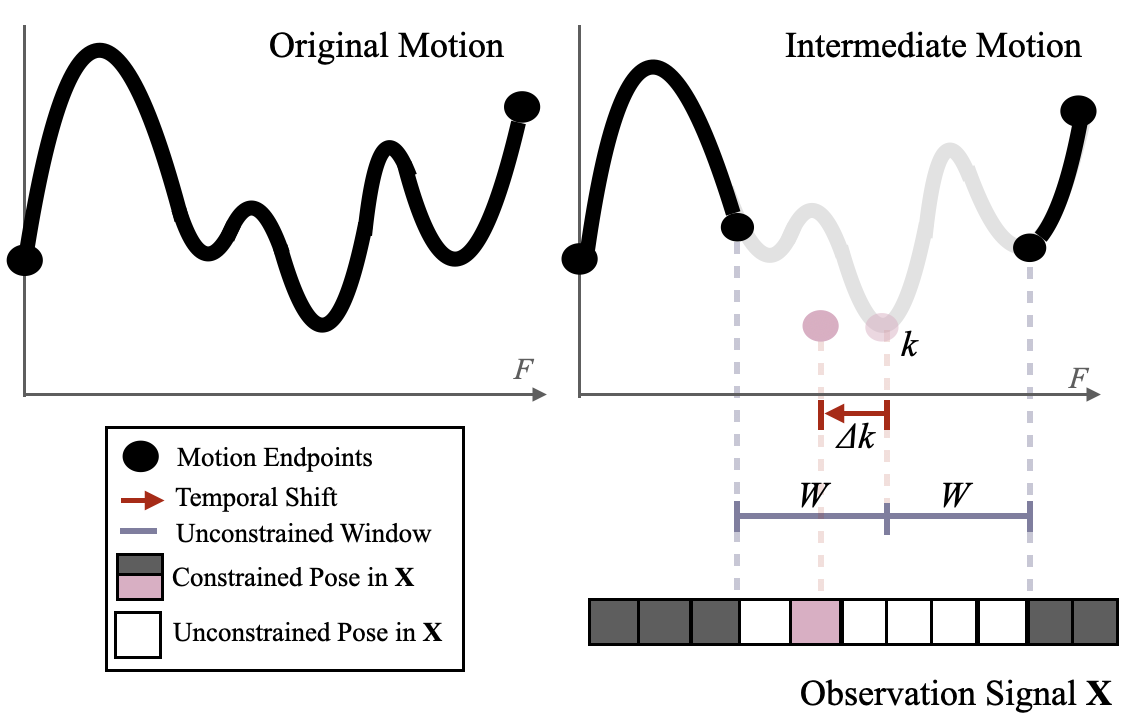}
    \caption{\textbf{Data Collection}: We synthetically generate plausible mistimed $\mathbf{X}$ from detailed motion clips $\mathbf{Y}$ (left, top). For each detailed motion sequence \detailed, we first identify poses that could plausibly have served as keyframes for \detailed. We select one at random and simulate \textit{approximate} timing by temporally shifting it by a small integer, which produces \blockingpose. We delete a window of neighboring frames (right, top). The submotions outside the deleted window, and \blockingpose, form observation signal \blocking (right, bottom).} 
    \label{fig:data_collection}
\end{figure}

While existing motion capture datasets contain hundreds of high-quality motion clips (plausible $\mathbf{Y}$ motions), they are not paired with plausible corresponding (and potentially mis-timed) keyframes. Our approach is to analyze clips from an existing dataset of high-quality motion $\mathbf{Y}$ to find frames that could have served as important keyframes for creating the clip. Then, we use these frames to synthetically generate $\mathbf{X}$ for each $\mathbf{Y}$.

\subsubsection{Selecting Keyposes} 

It is common for animators to place keyframes at extrema in the desired motion, such as the highest point of a jump~\cite{animatorssurvivalkit}. To model this animator behavior, we analyze detailed motion clips to identify poses marking extremes. Specifically, we find all poses with extrema in the character's root or end effectors and select one of these poses $\mathbf{X}_k$ at random to be a keyframe constraint.

\subsubsection{Approximate Timing } As we have described, a pose at ground truth frame index \textit{k} in \detailed, \detailedpose, may appear at a different frame in \blocking, i.e., \blockingpose, as in practice defining the exact timing and spacing for keyframe constraints can be challenging and/or tedious. To mimic how poses in $\mathbf{X}$ may not occur at precisely the same frame that they do in $\mathbf{Y}$, we temporally shift \detailedpose by a randomly selected frame delta $\Delta k \in [-P, P)$, where \textit{P} is a small integer, resulting in an imprecisely timed pose \blockingpose.

\subsubsection{Constructing $\mathbf{X}$}
\label{sec:gen_blocking_motion}

Consider two practical motion generation scenarios. In a synthesis task, an animator specifies a set of keyframes for a punching motion; in this case, $\mathbf{X}$ contains only those keyframes. In an editing task, the animator wants to modify an existing punch motion by making the character punch again. They identify frame index $k$ where the character’s arm is retracted after the first punch and specify a new keyframe for the desired articulation of a second punch. Here, $\mathbf{X}$ may contain some context from the original punch along with the new keyframe. Thus, $\mathbf{X}$ can include any amount of high-fidelity motion at inference time, depending on the specific requirements of the task (See Fig.~\ref{system}).

To mimic this trait in our dataset, we delete the detailed motion in the pose range $\mathbf{X}_{k - W:k+ \Delta k, k+ \Delta k +1:k + W }$, where $W \in [1, F)$ is a randomly chosen window of frames. $k \pm W$ is clamped between frame index $[1, F-2]$, and all frames outside this window are retained along with \blockingpose in $\mathbf{X}$. See Fig.~\ref{fig:data_collection} for an illustration.

When trained on a dataset containing randomly chosen $W$s, a model must understand which part of the $\mathbf{X}$ requires the most change, and how to add timing and spatial details in order to better match realistic motion.

\subsection{Model}

With a (synthetic) dataset of (\blocking, \detailed) pairs in hand, we use our generated training data to train a conditional diffusion model that turns approximately-timed keyframe constraints into high-fidelity, well-timed animation. Specifically, we seek to learn from data how to add precise timing and spatial details to $\mathbf{X}$ to create $\mathbf{Y}$. The transformation consists of a global time warp $\mathbf{w} \in \mathbb{R}^{F \times 1}$ and pose residuals $\Delta \mathbf{X} \in \mathbb{R}^{F \times D}$, where $D$ is the dimension of the pose representation.

\begin{equation}
 \mathbf{Y} = \texttt{warp}(\mathbf{w}, \mathbf{X}) + \Delta\mathbf{X}
\label{eq:warp}
\end{equation}

The warp operator \functionwarp uses the time warp function defined by \textbf{w} to modify the global timeline of \blocking. A time warp function can be thought of a mapping function that takes original, continuous time values of a motion sequence and transforms them into new time values. Thus given the original timeline \textit{T} and modified timeline $T'$, the time warp function $\mathbf{w}$ maps each original frame index to a new frame index:

\begin{equation} 
T '= \mathbf{w}(T) 
\end{equation}

\edit{We implement \textbf{w} as a backward mapping, e.g., $\mathbf{w}$ determines, for each warped time value, which original time value $T_{i}$ it corresponds to. Since $T_{i}$ may not align exactly with a discrete frame number, i.e., $T_{i}$ may be a non-integer value, we use bilinear interpolation to estimate the pose at non-integer frame values. $\mathbf{w}$ itself is parameterized as a $F$-dim vector, where the value at index $f$ represents the slope of $\mathbf{w}$ at frame $f$. A cumulative sum reconstructs $\mathbf{w}$ from this vector.}  Further implementation details can be found in the Supplement.

Recall that unconstrained regions in $\mathbf{X}$ are undefined. A smooth global time-warp of a motion comprising both defined and undefined regions is not well-defined because the transformation cannot be consistently extended to the undefined areas. Replacing unconstrained regions with noise or a scalar still introduces ambiguity regarding how the warp should behave in those regions. Moreover, it leads to abrupt changes between the constrained and unconstrained regions.

Instead, \edit{similar to the formulation in ~\cite{oreshkin2022motion, twostagetransformers}}, we replace poses in unconstrained regions of $\mathbf{X}$ with values obtained through linear spline interpolation between the boundaries of the unconstrained region before inputting $\mathbf{X}$ to the model. Our reasoning is that linearly interpolated motion preserves the constraints in $\mathbf{X}$, is a reasonable prior for coarse approximation of \detailed, can be automatically generated, and is defined everywhere in $[0, F)$.

Extracting retimed constraints following global time-warping, e.g., to then spatially infill them with an existing motion-inbetweening model, is not easily differentiable. Instead, we represent spatial detail as pose residuals \residual, which are added to the warped \blocking. We use our synthetic dataset to learn both $\mathbf{w}$ and \residual.

\subsubsection{Diffusion Model} 
We leverage a diffusion model to learn $\mathbf{w}$ and \residual. The core component in diffusion models is a denoising network, $U$. Given a forward Markov noising process and ground-truth motion $\mathbf{Y}$,
\begin{equation}
q(\mathbf{Y}^t | \mathbf{Y}) = \mathcal{N}(\sqrt{\alpha_t}\mathbf{Y}, (1-\alpha_t)I)
\end{equation}
where $\alpha_t \in(0, 1) $ is a constant that decrease monotonically with $t$, $U$ is trained to reverse the forward diffusion process. In our setup, \textit{U} acts as a denoiser that takes as input the noised target motion $\mathbf{Y}^{t}$, the condition \blocking, and the current diffusion step $t$, and outputs a prediction of the denoised motion \detailed with objective:
\begin{equation}
\mathcal{L} = \mathbf{E}_{\mathbf{Y}, t} [ \| \mathbf{Y} - U(\mathbf{Y}^{t}, \mathbf{X}, t) \|^2_2 ].
\label{eq:rec_loss}
\end{equation}



\begin{figure}
    \centering
    \includegraphics[width=0.9\linewidth]{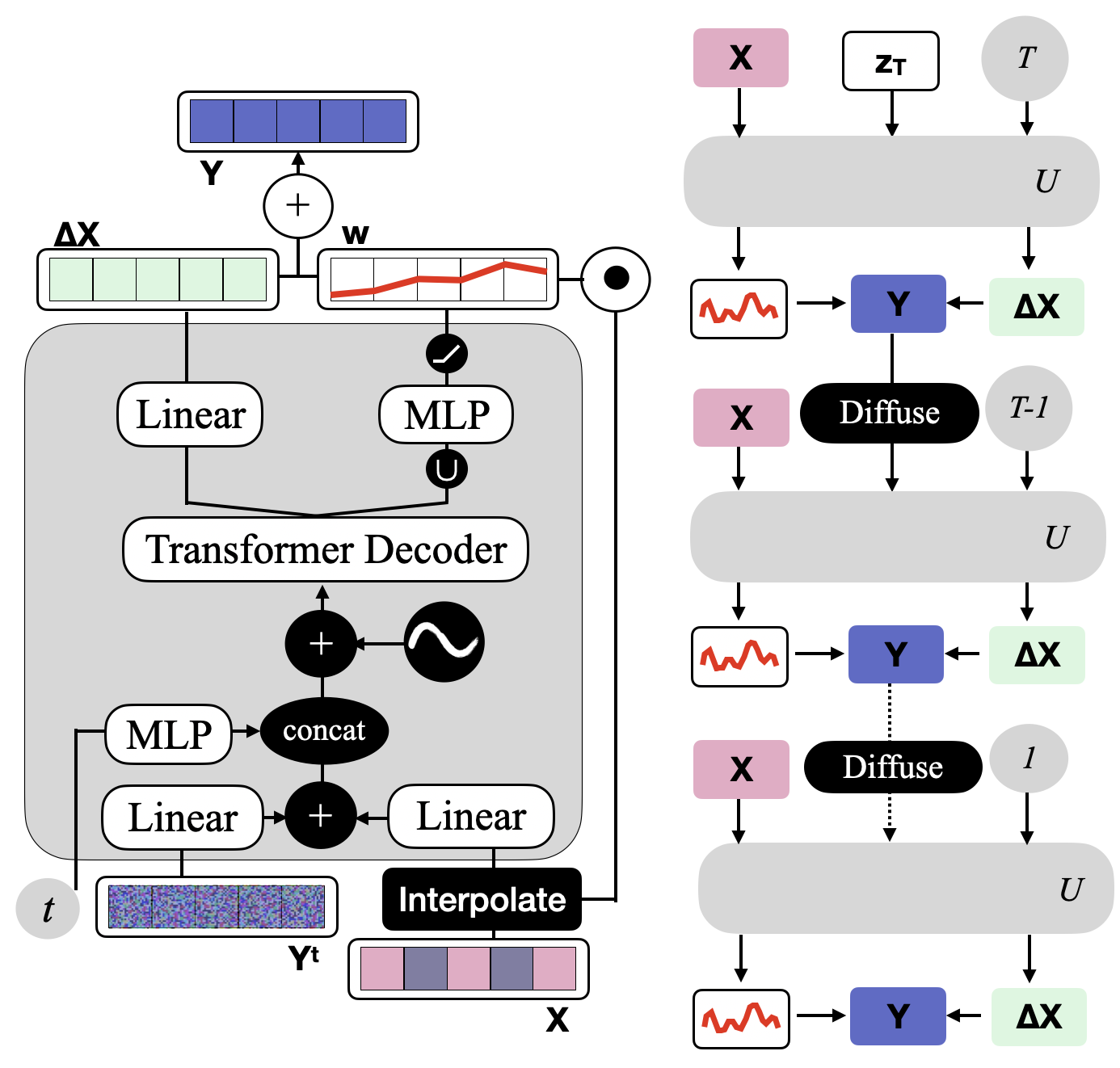} \\
    \caption{\textbf{Diffusion Model Architecture}: During training, our two-headed model \textit{U} (left) learns to predict both a time warp $\mathbf{w}$ and pose details \residual from a shared transformer decoder backbone, given observation signal \blocking (preprocessed so that all undefined regions are replaced with an interpolation solution), diffusion timestep $t$, and a noisy sequence $\mathbf{Y}^{t}$. $\mathbf{w}$ is applied to \blocking as a global retiming operation, then summed with \residual as a pose detailing operation. At inference (right), \textit{U} iteratively denoises  the sequence from $t=T$ to $t=0$. We use $\cup$ to represent the ``flatten'' operator. } 
    \label{fig:model}
\end{figure}

\noindent
\paragraph{\textbf{Network Architecture}}
Our model consists of a two-headed network architecture that incorporates a transformer decoder, illustrated in Figure ~\ref{fig:model}. \blocking, the condition signal, is first preprocessed to infill unconstrained regions with an interpolation solution, as described above, then projected to the transformer dimension ~\textit{L} via a feed-forward network, as is diffusion timestep~\textit{t}. The resulting encodings are concatenated and summed with a positional embedding; projected frames are then fed into the transformer decoder. Separately, the projected \blocking also acts as external memory to the transformer decoder, which outputs a per-frame latent representation. Finally, two separate heads process the decoder output. A pose residual branch projects the result into the original motion dimension, producing $\Delta \mathbf{X}$. Separately, a warping branch processes the decoder output, converting the encoding of individual frames into a representation of the entire motion by flattening the frame dimension, i.e., $\mathbb{R}^{B \times D \times F} \rightarrow \mathbb{R}^{B \times DF}$, where \textit{B} is the batch size and \textit{D} is the dimension of the pose representation. The new encoding is processed via a 3-layer MLP and a final prelu activation, which results in the predicted $\mathbf{w}$.

Because the reconstruction loss in Equation~\ref{eq:rec_loss} is calculated subject to Equation~\ref{eq:warp}, gradients are propagated through both heads into the shared backbone. We only calculate the reconstruction loss on the predicted $\mathbf{Y}$, not on the predicted $\mathbf{w}$ or $\Delta \mathbf{X}$.

\paragraph{ \textbf{Long Motion Conditions}} Though our model is trained to produce motions of fixed length \textit{F}, we extend the approach of ~\cite{tseng2022edge} to handle arbitrary length \blocking at inference. First, we preprocess \blocking by splicing it into a batch of subsequences of length $F$. We transform the subsequences such that the first half of each subsequence matches the last half of the previous subsequence. We similarly constrain the intermediate predictions at each denoising step during inference, such that predicted subsequences can be concatenated into a single \detailed with the desired motion length. We direct readers to~\cite{tseng2022edge} for further details. 

\paragraph{\textbf{Inference-time Workflow}}

\edit{At inference, the set of input constraints \textit{K} contains user-provided keyframes (perhaps imprecisely-timed) and, in editing workflows, any portion of the original motion that the user wishes to retain. These constraints are placed on the timeline, forming $\mathbf{X}$ (see  Fig.~\ref{system}). Unconstrained regions of $\mathbf{X}$ are filled using linear interpolation. The model then generates $\mathbf{Y}$, conditioned on $\mathbf{X}$, adding pose and timing detail over \textit{T} diffusion denoising steps (see Fig.~\ref{fig:model}). }

\subsection{Implementation Details}
\subsubsection{Motion Representation} Each motion $\mathbf{X}$ and $\mathbf{Y}$ is represented as a sequence of poses in the SMPL format~\cite{SMPL:2015}. For pose state at frame \textit{f}, we represent each of the 24 joint angles using a 6D continuous representation~\cite{zhou2020continuity}. Each pose also contains a single 3-dim global translation, and a shape parameter $\beta \in \mathbb{R}^{10}$. We use a binary label for the heel and toe of both feet to represent contact with the ground, $\mathbf{b} \in \{0, 1\}$. We set the foot contact label to 0 for the condition $\mathbf{X}$, as foot contact is unknown in unconstrained regions. We also include the global joint positions as a redundant representation. The final pose representation is $\mathbf{Y}_f \in \mathbb{R}^{236}$, and a motion comprising \textit{F} frames is therefore $\mathbf{Y} \in \mathbb{R}^{F \times 236}$.

\subsubsection{Dataset Generation and Diffusion Model}
In this work, we use the motions in popular human motion capture dataset HumanML3D~\cite{AMASS:ICCV:2019}. We use motion clip length $F=60$, though we can support arbirary motion lengths (see Sec. 4.2.1.1). We use $P$=5 in our dataset generation process when temporally shifting keyframes. We train diffusion model \textit{U} using a NVIDIA Tesla V100 GPU for about 24 hours, using a batch size of 64; hyperparameters can be found in the Supplement. We run the inference process with $T=1000$ diffusion steps, which takes about 30 seconds for a batch size of 50 on a single GPU. The output may then be optionally post-processed with, e.g., foot-skate clean-up. 

\begin{figure*}
\centerline{
\setlength\tabcolsep{2pt}
\renewcommand{\arraystretch}{2.0}
\begin{tabular}{cccccccc}
\multicolumn{8}{c}{Observation Signal \blocking} \\
\includegraphics[width=0.12\linewidth, trim={8cm 6cm 8cm 4cm}, clip]{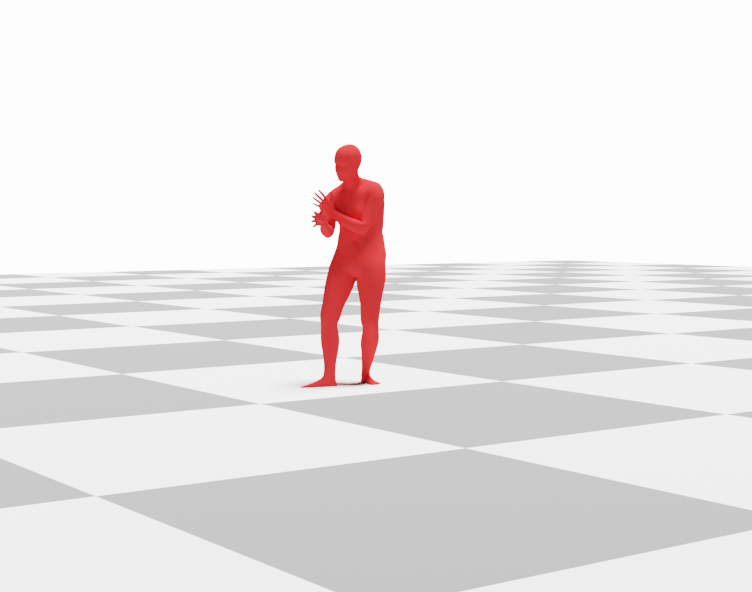} &
\includegraphics[width=0.12\linewidth, trim={15cm 6cm 1cm 4cm}, clip]{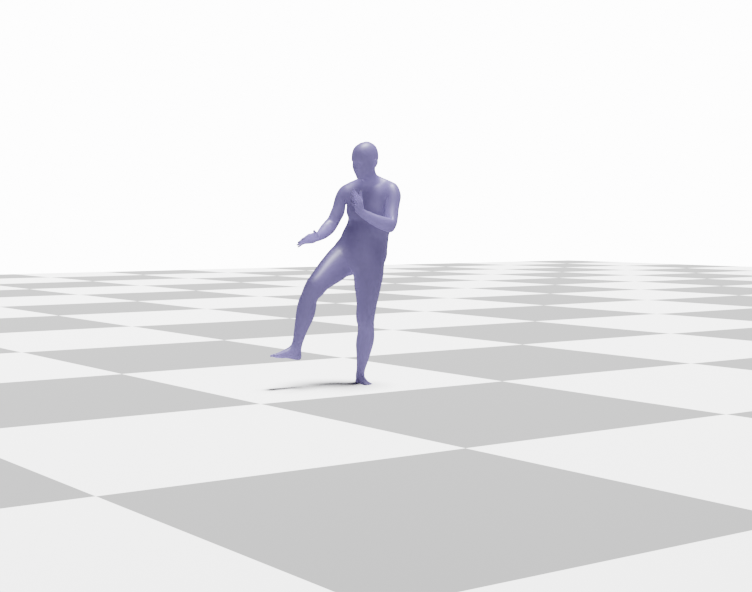} &
\includegraphics[width=0.12\linewidth, trim={15cm 6cm 1cm 4cm}, clip]{figures/synthesis/original/00005.png} &
\includegraphics[width=0.12\linewidth, trim={8cm 6cm 8cm 4cm}, clip]{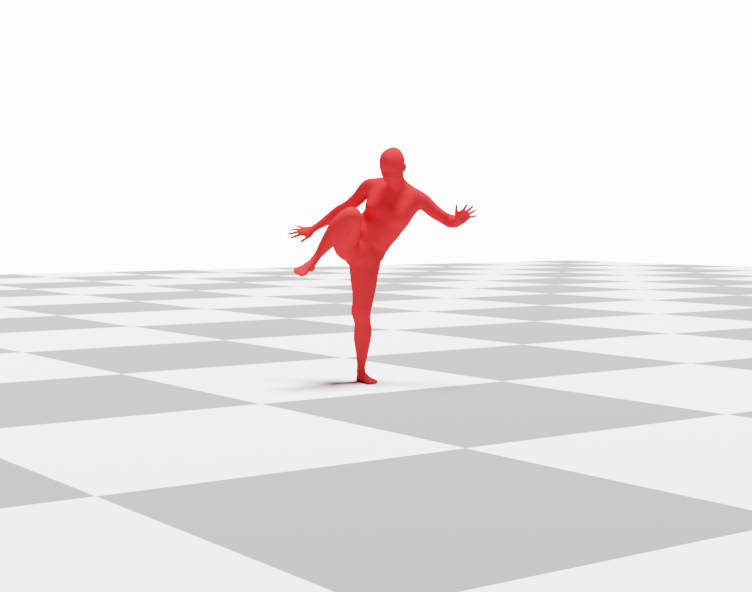} &
\includegraphics[width=0.12\linewidth, trim={15cm 6cm 1cm 4cm}, clip]{figures/synthesis/original/00005.png} &
\includegraphics[width=0.12\linewidth, trim={15cm 6cm 1cm 4cm}, clip]{figures/synthesis/original/00005.png} &
\includegraphics[width=0.12\linewidth, trim={15cm 6cm 1cm 4cm}, clip]{figures/synthesis/original/00005.png} &
\includegraphics[width=0.12\linewidth, trim={8cm 6cm 8cm 4cm}, clip]{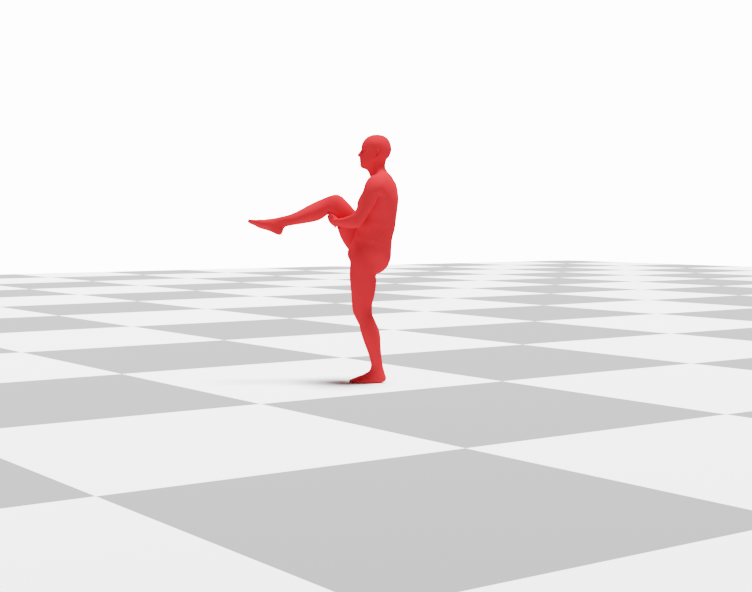} \\
\multicolumn{8}{c}{Generated \detailed} \\
\includegraphics[width=0.12\linewidth, trim={8cm 6cm 8cm 4cm}, clip]{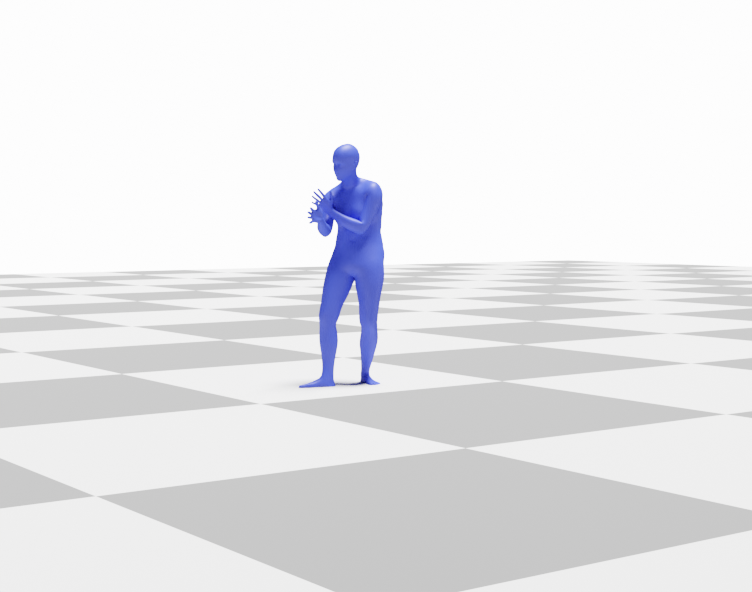} &
\includegraphics[width=0.12\linewidth, trim={8cm 6cm 8cm 4cm}, clip]{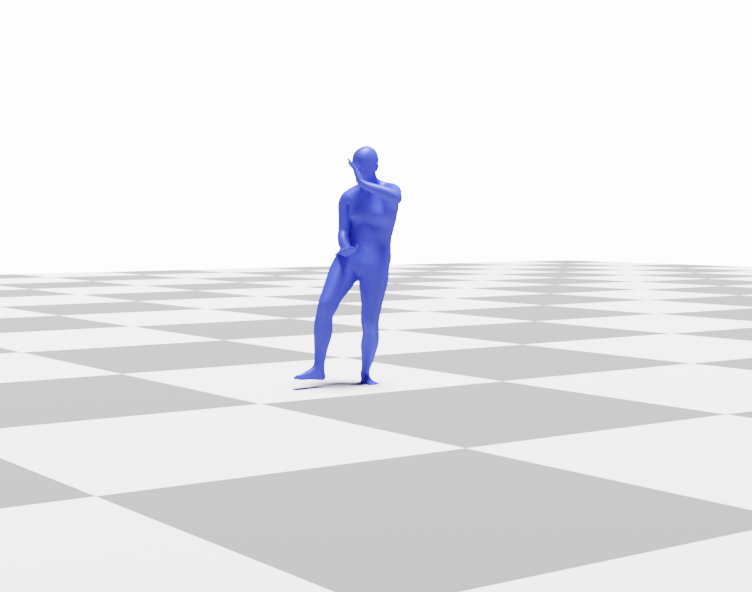} &
\includegraphics[width=0.12\linewidth, trim={8cm 6cm 8cm 4cm}, clip]{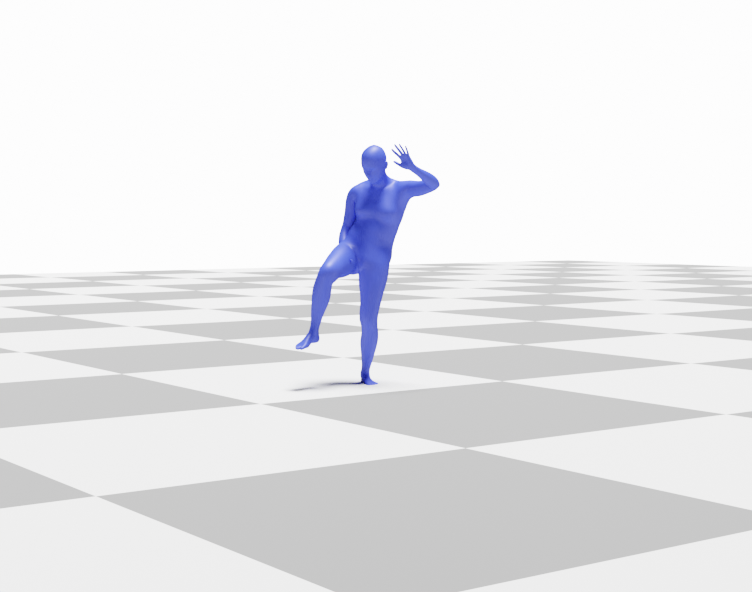} &
\includegraphics[width=0.12\linewidth, trim={8cm 6cm 8cm 4cm}, clip]{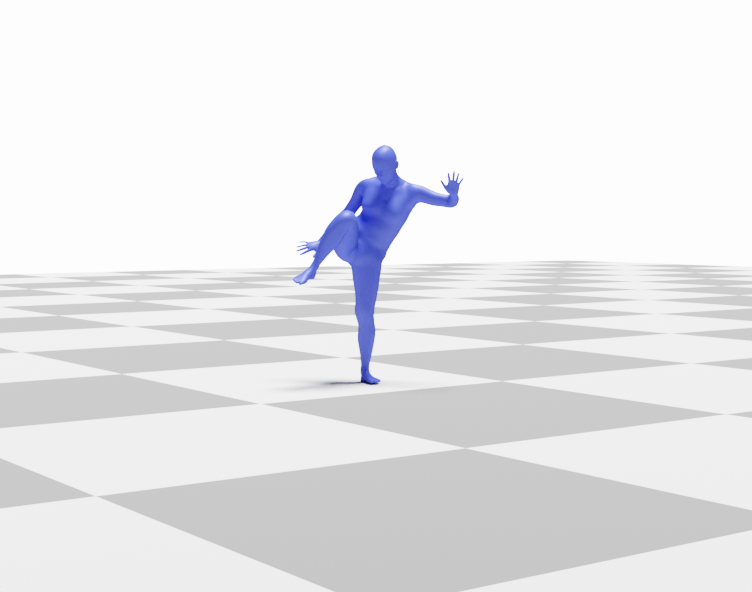} &
\includegraphics[width=0.12\linewidth, trim={8cm 6cm 8cm 4cm}, clip]{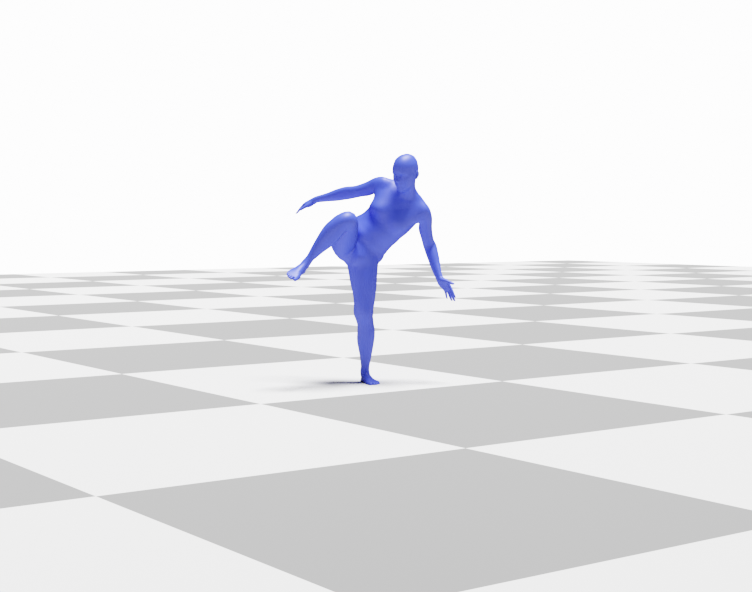} &
\includegraphics[width=0.12\linewidth, trim={8cm 6cm 8cm 4cm}, clip]{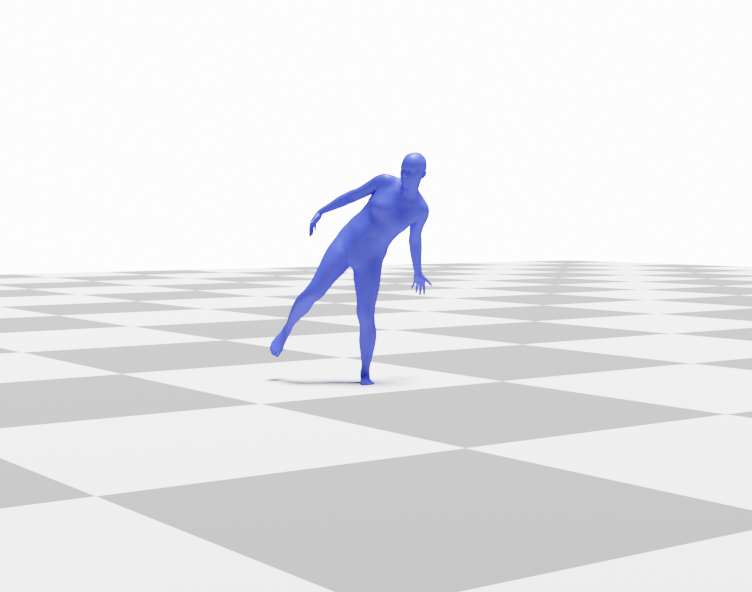} &
\includegraphics[width=0.12\linewidth, trim={8cm 6cm 8cm 4cm}, clip]{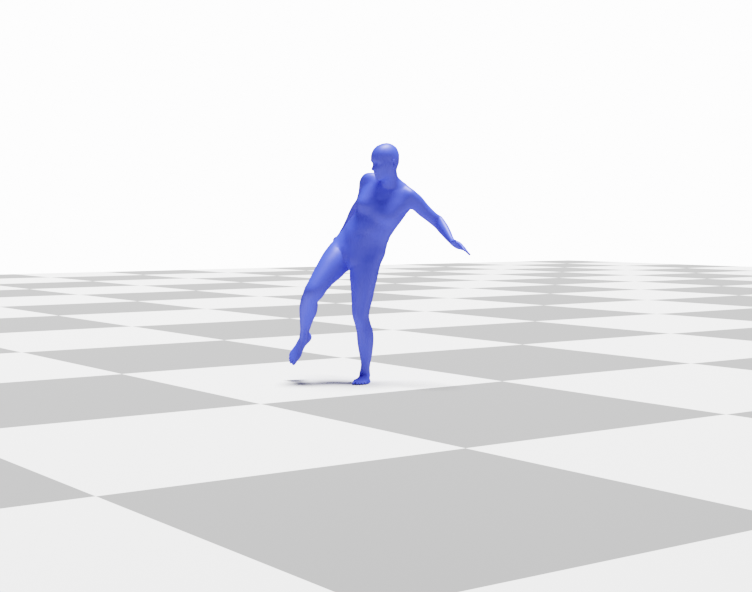} &
\includegraphics[width=0.12\linewidth, trim={8cm 6cm 8cm 4cm}, clip]{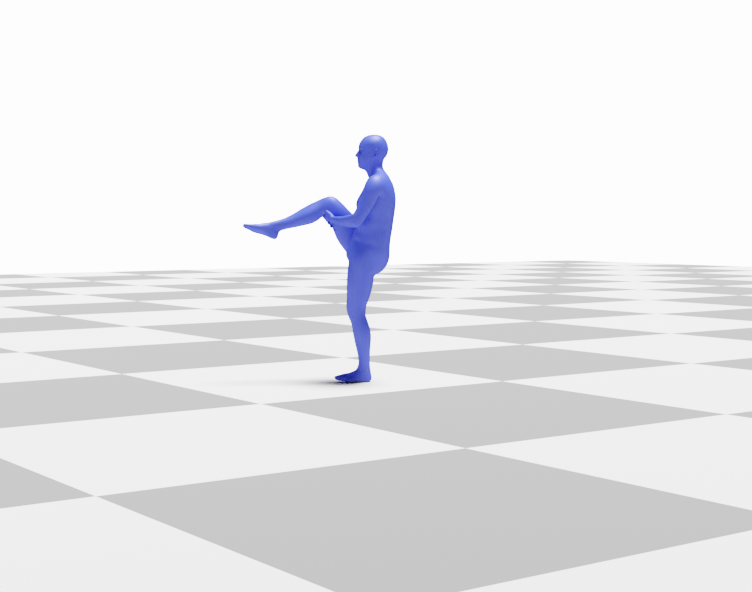} \\
\multicolumn{8}{c}{Generated \detailed, different seed} \\
\includegraphics[width=0.12\linewidth, trim={8cm 6cm 8cm 4cm}, clip]{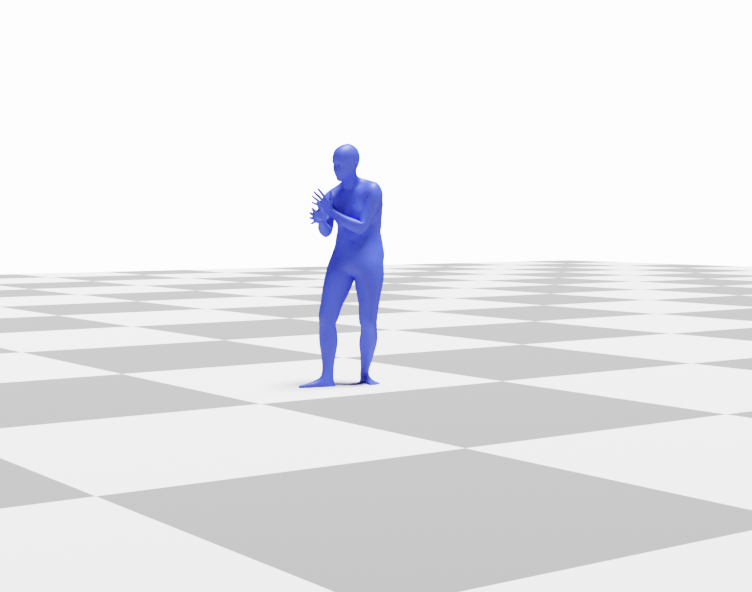} &
\includegraphics[width=0.12\linewidth, trim={8cm 6cm 8cm 4cm}, clip]{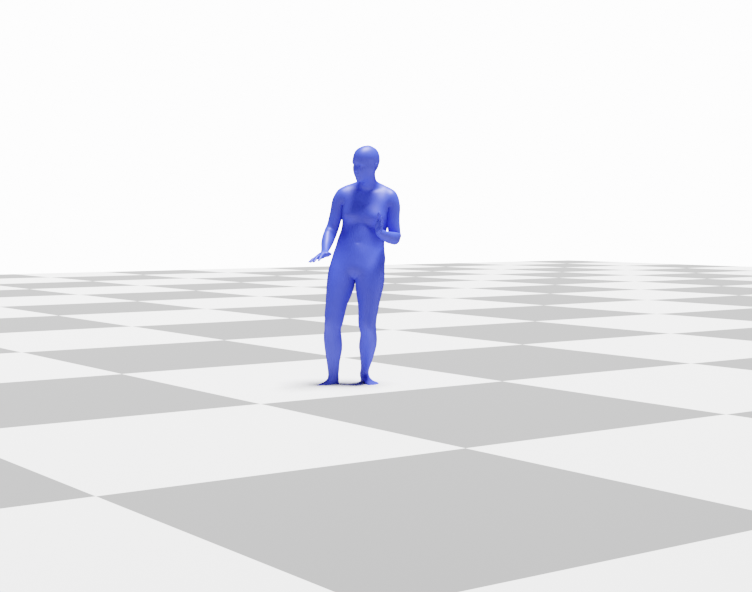} &
\includegraphics[width=0.12\linewidth, trim={8cm 6cm 8cm 4cm}, clip]{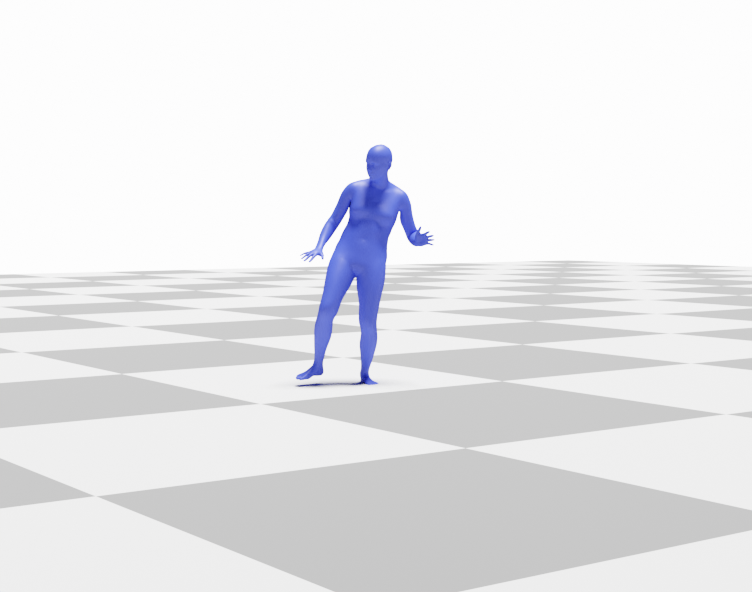} &
\includegraphics[width=0.12\linewidth, trim={8cm 6cm 8cm 4cm}, clip]{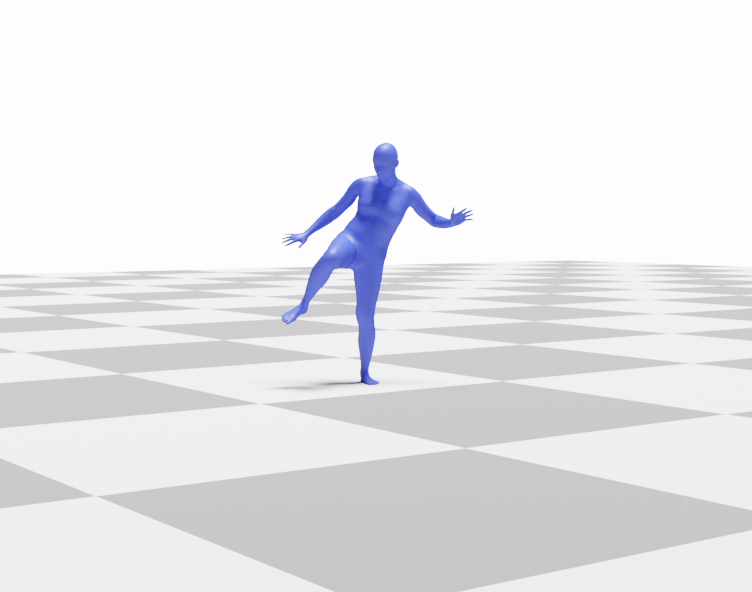} &
\includegraphics[width=0.12\linewidth, trim={8cm 6cm 8cm 4cm}, clip]{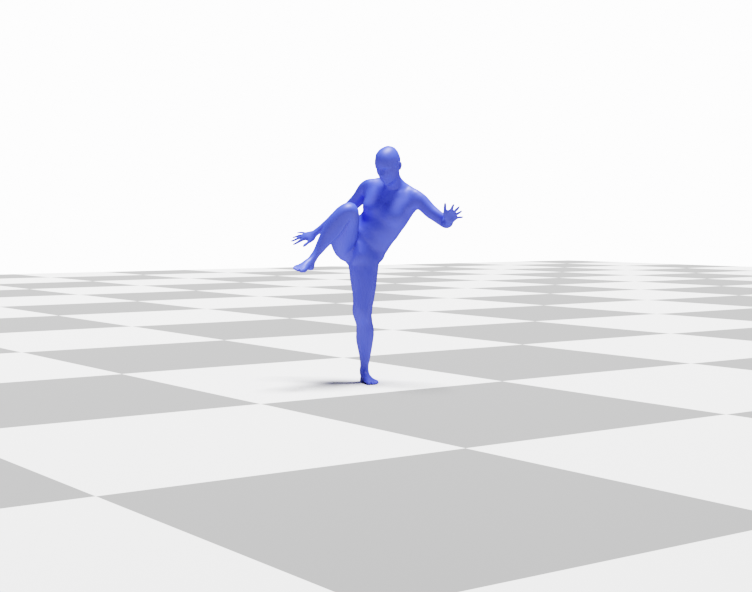} &
\includegraphics[width=0.12\linewidth, trim={8cm 6cm 8cm 4cm}, clip]{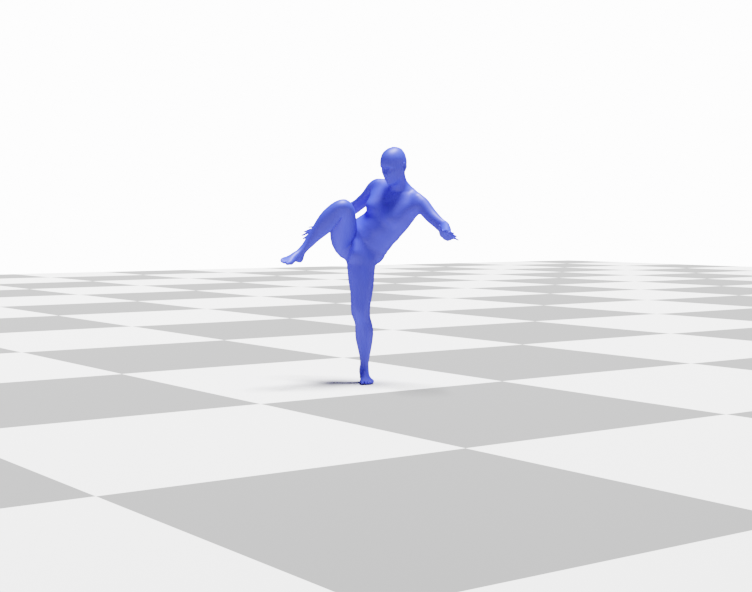} &
\includegraphics[width=0.12\linewidth, trim={8cm 6cm 8cm 4cm}, clip]{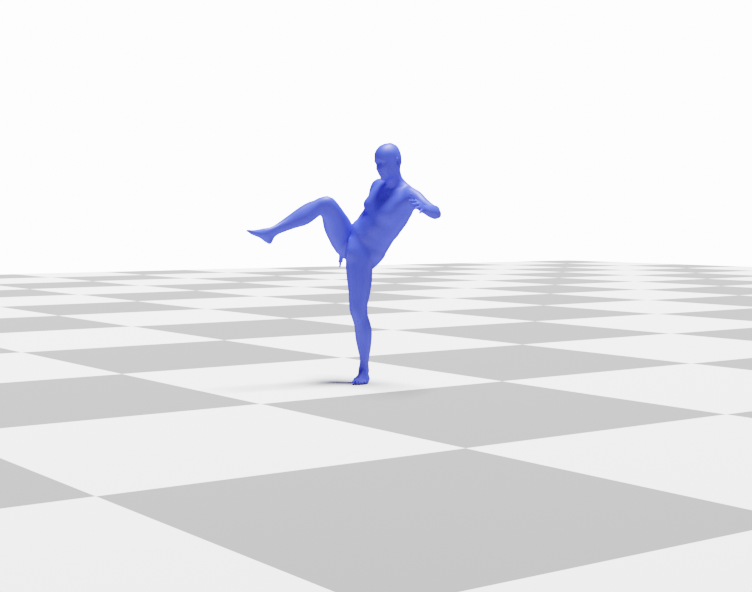} &
\includegraphics[width=0.12\linewidth, trim={8cm 6cm 8cm 4cm}, clip]{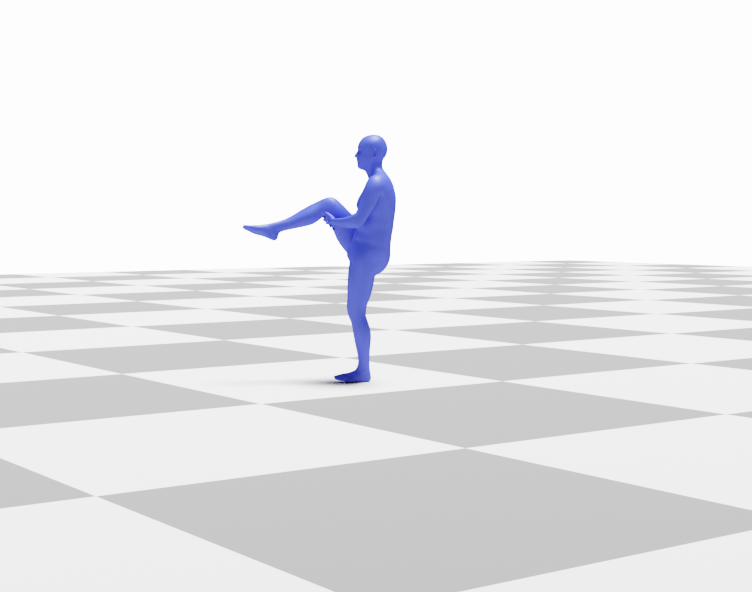} \\
\end{tabular}
}
\caption{ \textbf{Motion synthesis:} starting from approximately timed keyframe constraints (top row, red) of a character raising and grabbing its right leg, our model generates detailed motion \detailed; we show two generations here (middle row, bottom row). Our model can capture different modes of motion with different seeds. One seed produces \detailed (middle row) where the character loses its balance, then recovers. Another seed produces \detailed (bottom row) where the character expertly grabs its leg and pivots. In the latter case, notice how the middle keyframe (top row, second red pose) appears a little later in the generated motion (bottom row).}
\label{fig:synthesis}
\end{figure*}
\section{Results}

The goal of our method is to convert approximately timed keyframe constraints $\mathbf{X}$  into well-timed, detailed motion. \detailed should therefore be similar in structure and dynamics to intended outcome, and contain the pose constraints in $\mathbf{X}$.

\subsection{Qualitative Evaluation}

We used our system to transform various, approximately-timed keyframe constraints into realistic, well-timed motions. These constraints were created in multiple ways: (a) by sampling and temporally perturbing keyframes from test set motions, and (b) by simulating typical user interactions with the system. For example, given a desired motion such as "a character ducks, then kicks, then ducks again," keyposes can be, e.g., selected and composed from different motions in the mocap data, formed by spatially combining upper/lower body poses, and extracted from images using 3D pose estimation methods ~\cite{li2021hybrik}. 

\begin{figure*}
\centerline{
\setlength\tabcolsep{2pt}
\renewcommand{\arraystretch}{2.0}
\begin{tabular}{cccccccc}
\multicolumn{8}{c}{Original Motion} \\
\includegraphics[width=0.12\linewidth, trim={7cm 6cm 9cm 4cm}, clip]{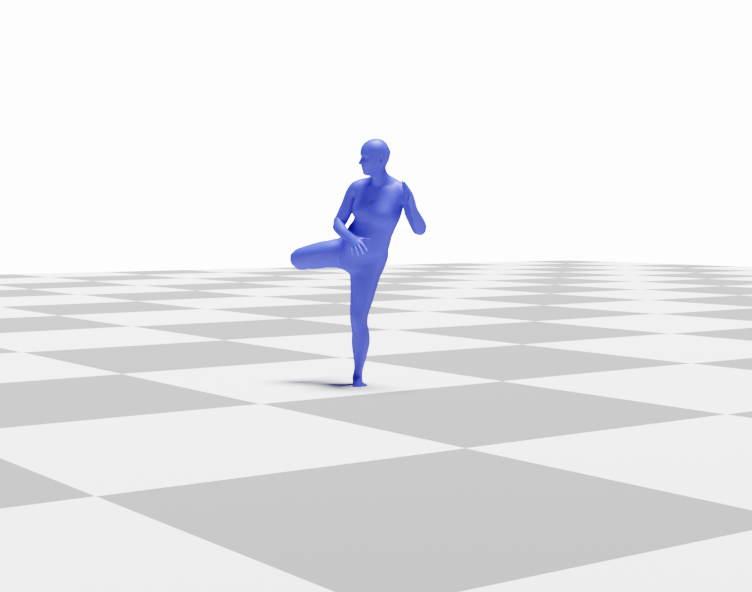} &
\includegraphics[width=0.12\linewidth, trim={7cm 6cm 9cm 4cm}, clip]{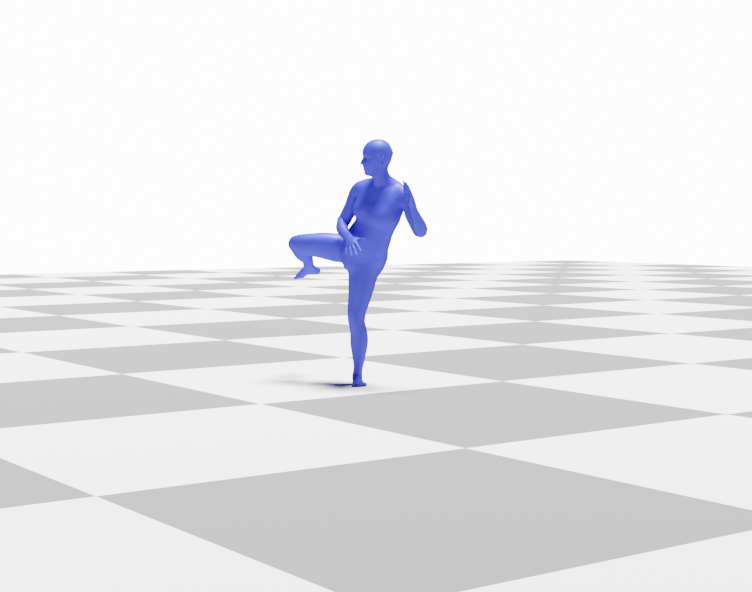} &
\includegraphics[width=0.12\linewidth, trim={7cm 6cm 9cm 4cm}, clip]{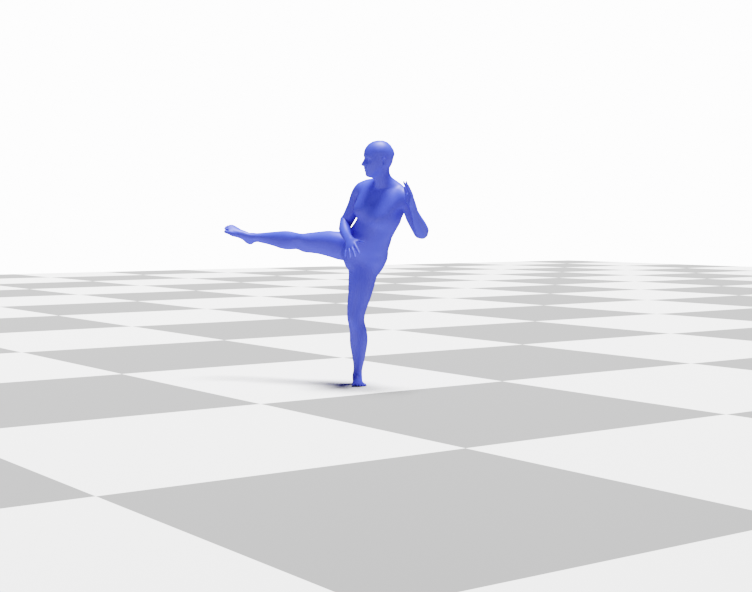} &
\includegraphics[width=0.12\linewidth, trim={7cm 6cm 9cm 4cm}, clip]{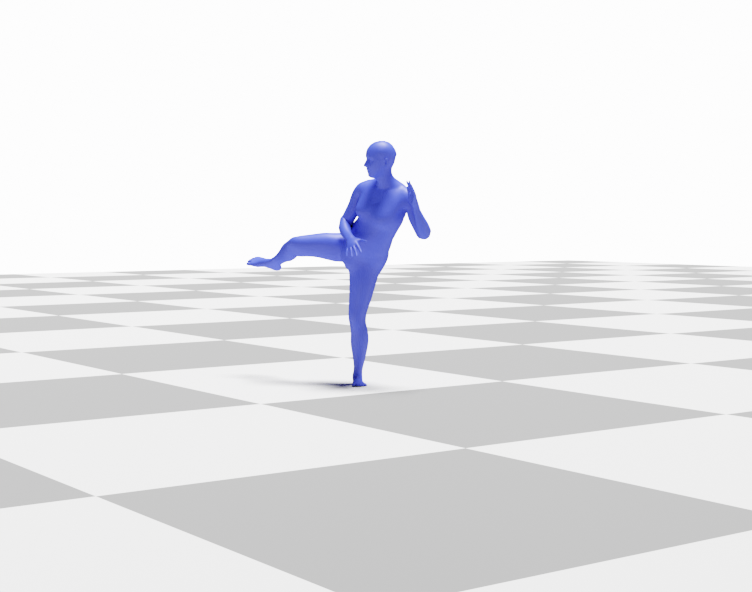} &
\includegraphics[width=0.12\linewidth, trim={7cm 6cm 9cm 4cm}, clip]{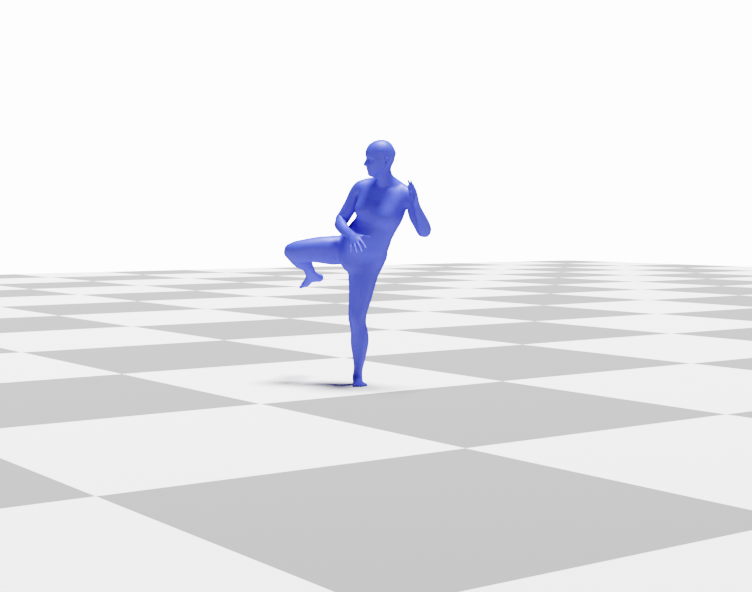} &
\includegraphics[width=0.12\linewidth, trim={7cm 6cm 9cm 4cm}, clip]{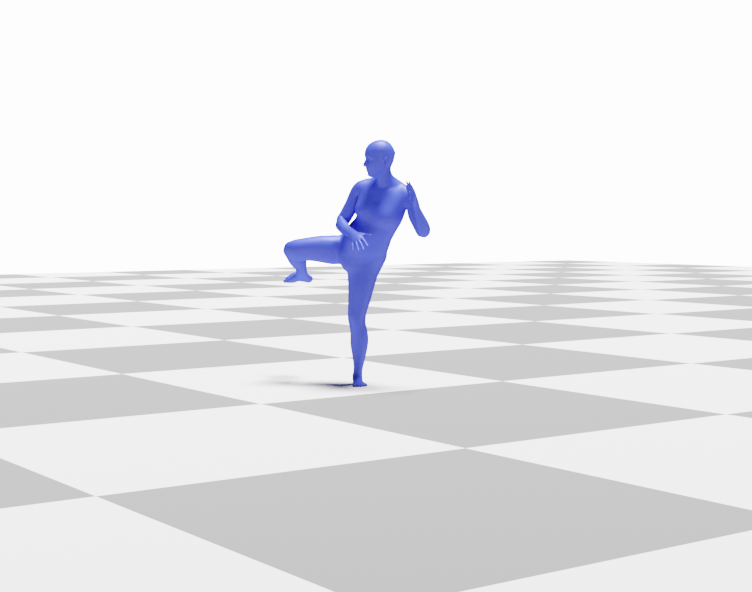} &
\includegraphics[width=0.12\linewidth, trim={7cm 6cm 9cm 4cm}, clip]{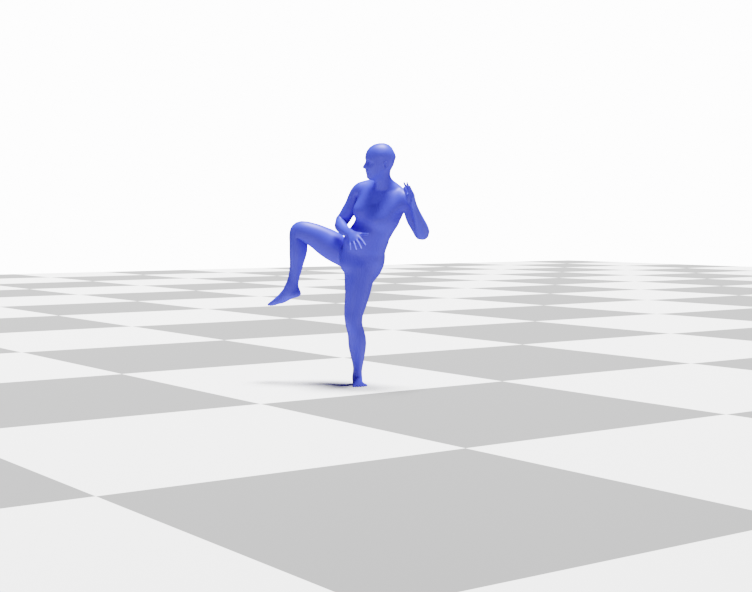} &
\includegraphics[width=0.12\linewidth, trim={7cm 6cm 9cm 4cm}, clip]{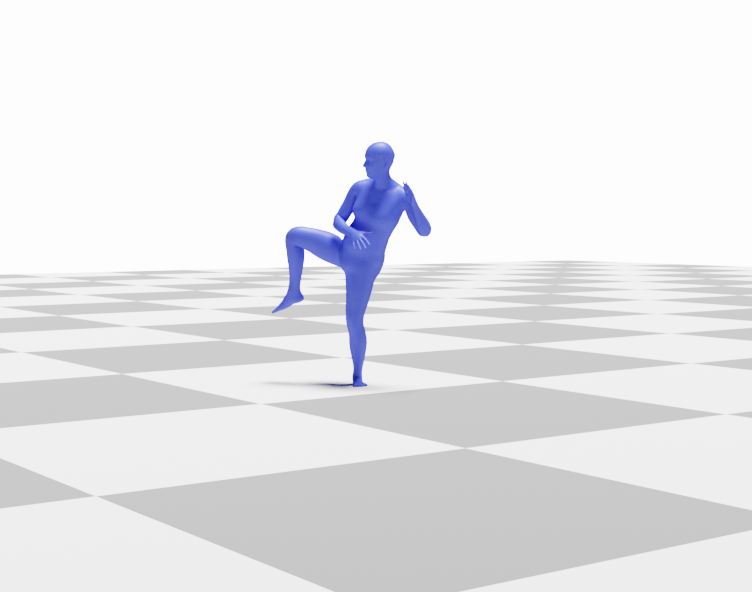} \\
\multicolumn{8}{c}{Observation Signal \blocking} \\
\includegraphics[width=0.12\linewidth, trim={7cm 6cm 9cm 4cm}, clip]{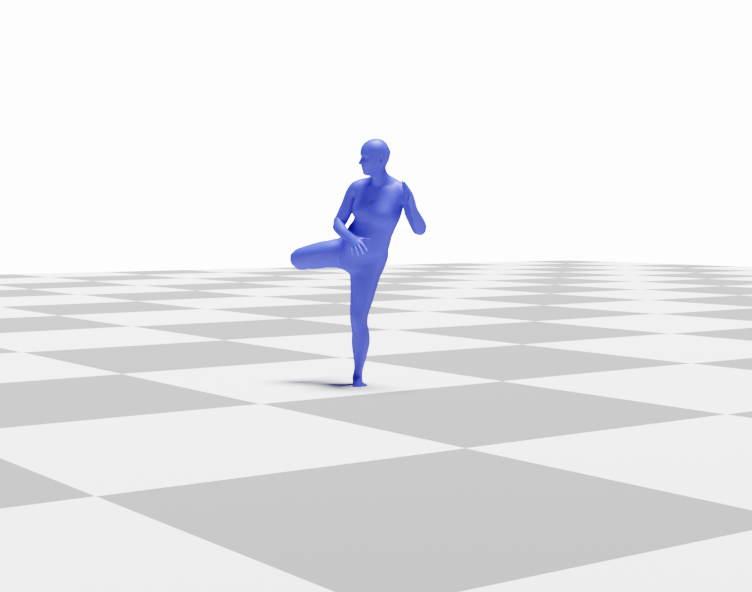} &
\includegraphics[width=0.12\linewidth, trim={7cm 6cm 9cm 4cm}, clip]{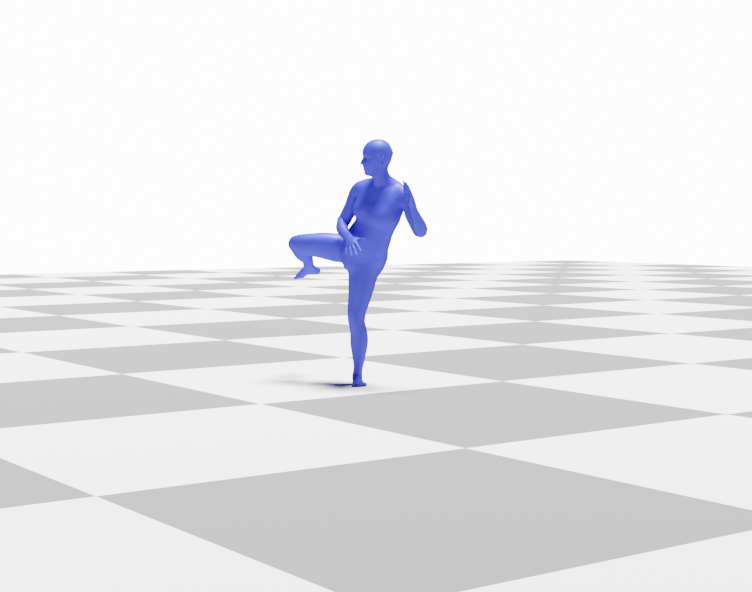} &
\includegraphics[width=0.12\linewidth, trim={7cm 6cm 9cm 4cm}, clip]{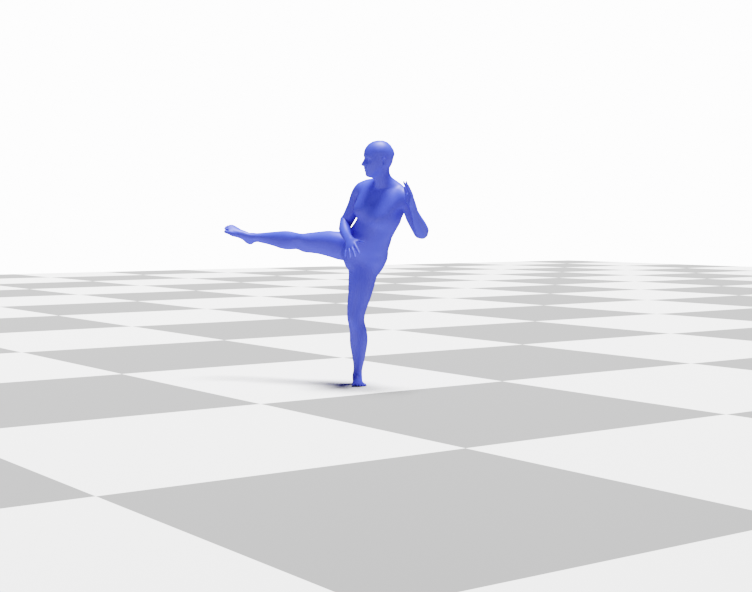} &
\includegraphics[width=0.12\linewidth, trim={16cm 6cm 0cm 4cm}, clip]{figures/editing/original/00007.png} &
\includegraphics[width=0.12\linewidth, trim={16cm 6cm 0cm 4cm}, clip]{figures/editing/original/00007.png} &
\includegraphics[width=0.12\linewidth, trim={7cm 6cm 9cm 4cm}, clip]{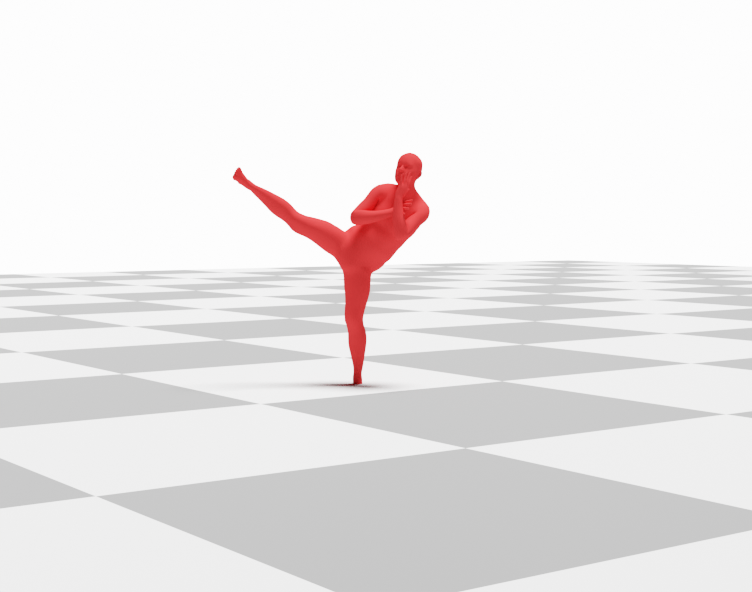} &
\includegraphics[width=0.12\linewidth, trim={16cm 6cm 0cm 4cm}, clip]{figures/editing/original/00007.png} &
\includegraphics[width=0.12\linewidth, trim={7cm 6cm 9cm 4cm}, clip]{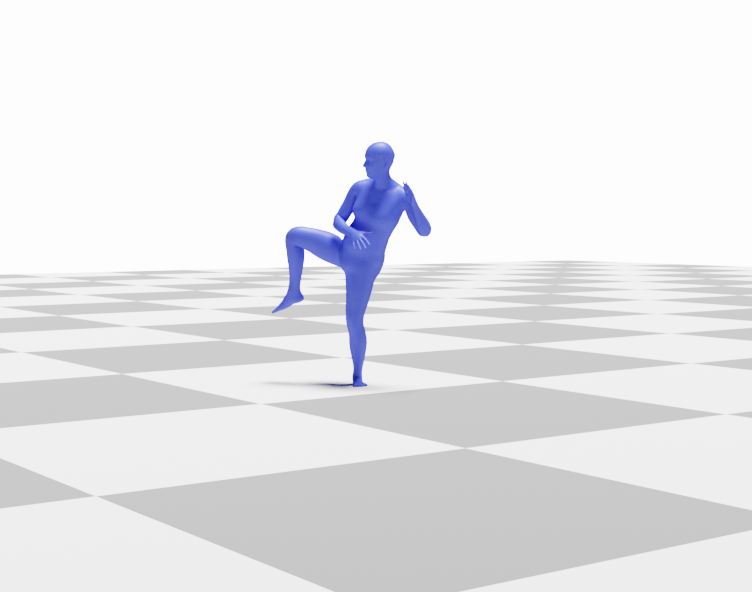} \\
\multicolumn{8}{c}{Generated \detailed} \\
\includegraphics[width=0.12\linewidth, trim={7cm 6cm 9cm 4cm}, clip]{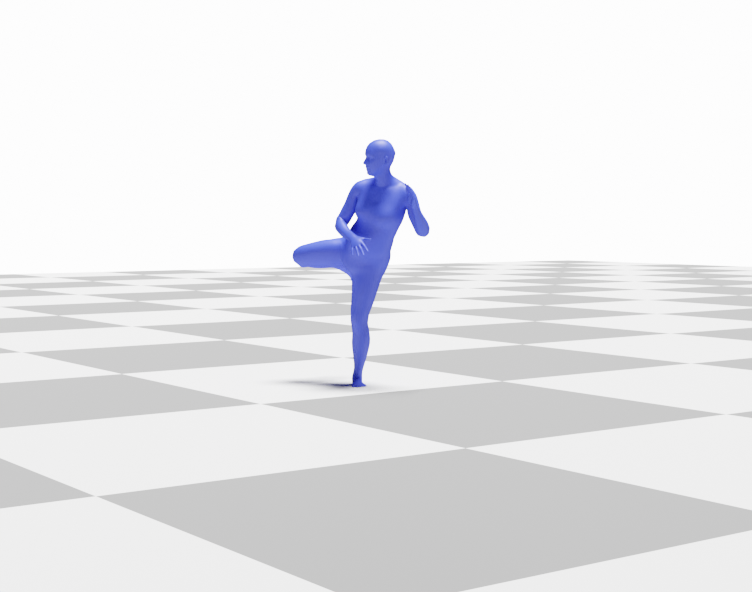} &
\includegraphics[width=0.12\linewidth, trim={7cm 6cm 9cm 4cm}, clip]{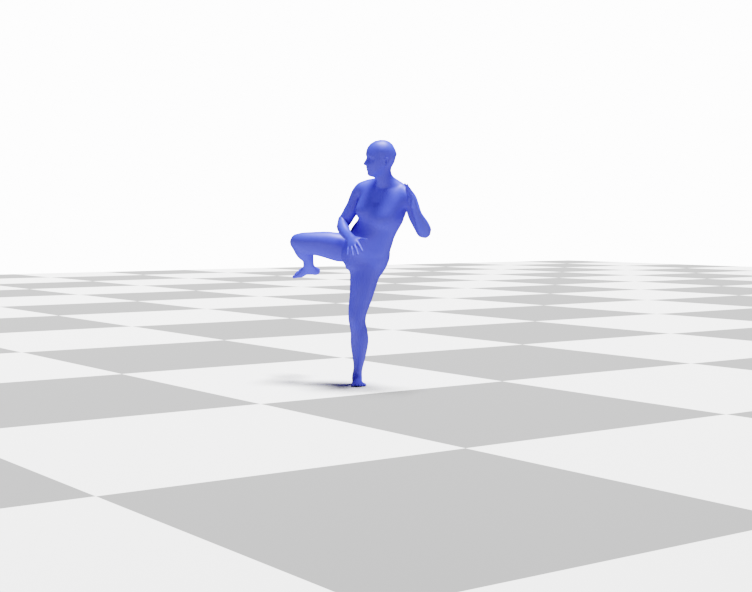} &
\includegraphics[width=0.12\linewidth, trim={7cm 6cm 9cm 4cm}, clip]{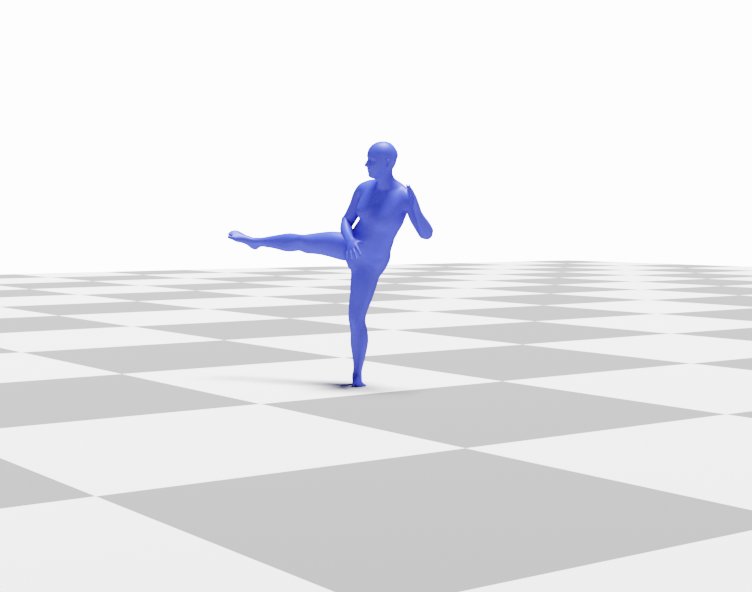} &
\includegraphics[width=0.12\linewidth, trim={7cm 6cm 9cm 4cm}, clip]{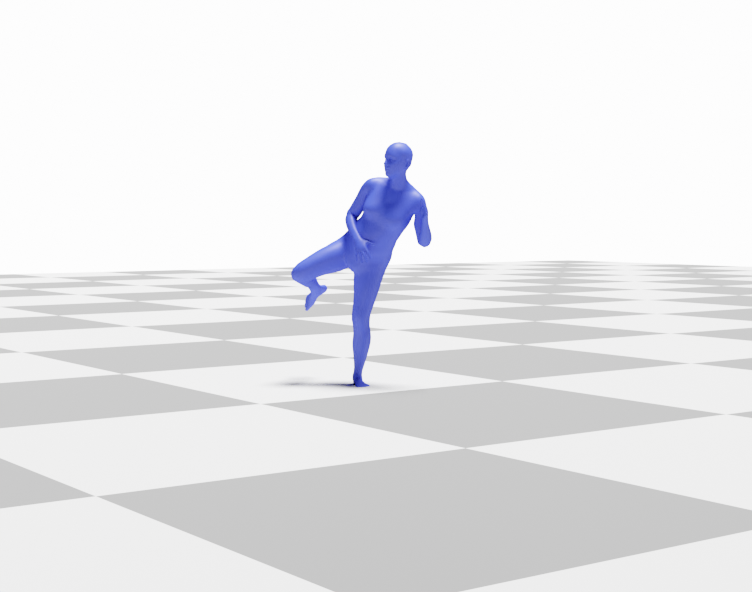} &
\includegraphics[width=0.12\linewidth, trim={7cm 6cm 9cm 4cm}, clip]{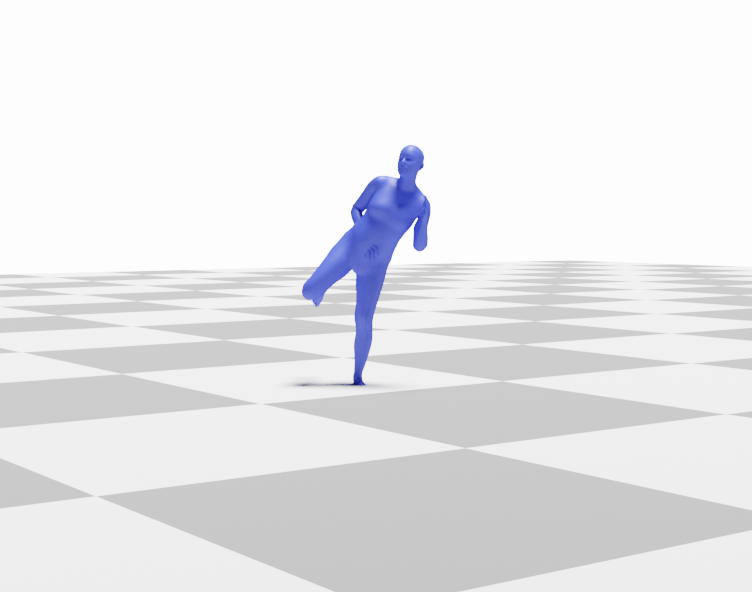} &
\includegraphics[width=0.12\linewidth, trim={7cm 6cm 9cm 4cm}, clip]{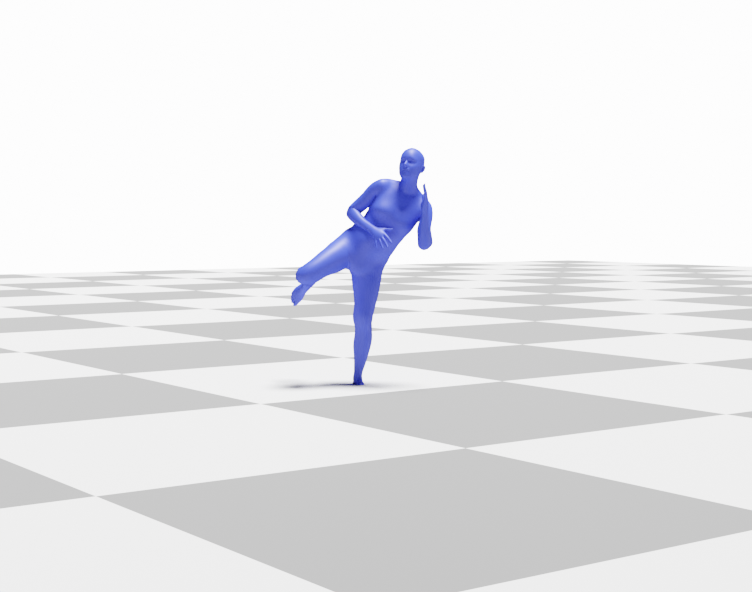} &
\includegraphics[width=0.12\linewidth, trim={7cm 6cm 9cm 4cm}, clip]{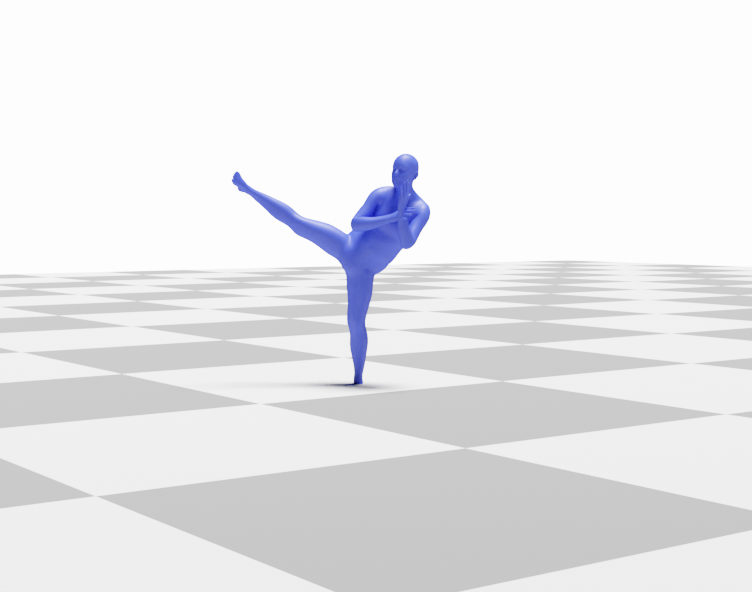} &
\includegraphics[width=0.12\linewidth, trim={7cm 6cm 9cm 4cm}, clip]{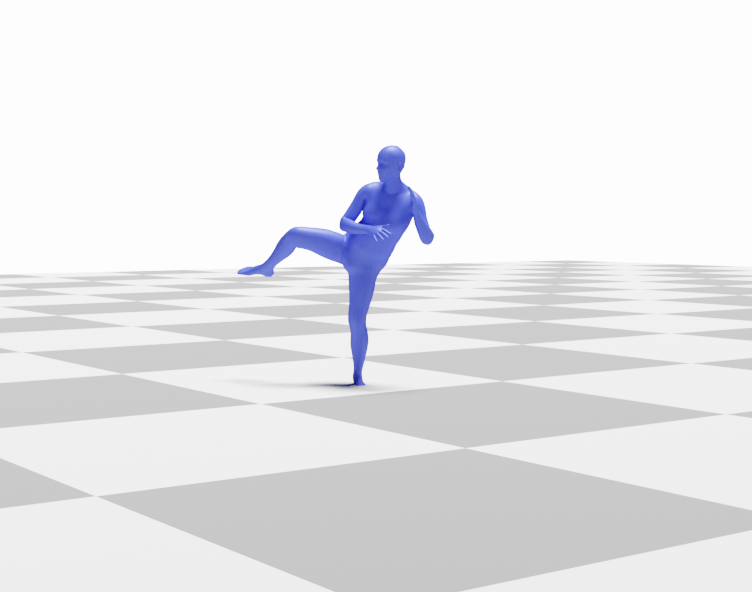} \\
\end{tabular}
}
\caption{\textbf{Motion Editing}: Given an existing motion (top) of a character kicking with the right leg, the animator wants to make the character kick a second time. The animator creates a new pose (middle row, red) where the character kicks again, and places it at approximately the right place on the timeline: twenty frames after the first kick. Given this input, our model generates detailed motion \detailed (bottom).  }
\label{fig:editing}
\end{figure*}

In Figures \ref{fig:banner}, \ref{fig:synthesis}, \ref{fig:editing}, our video, and Supplemental Materials, we demonstrate how the method can generate detailed motion with plausible timings, in diverse subjects like martial arts, dance, and navigation. These also illustrate how the model generalizes to be able to handle different numbers and frame placements of keyframe constraints. Fig~\ref{fig:synthesis} also shows how different generation seeds can capture different motion behavior modes.

\subsection{Quantitative Evaluation}

\subsubsection{Experimental Set-up} To quantitatively evaluate our method, \textbf{LT} (loose-timing), we compare performance against three baselines that test the importance of our dataset generation scheme and our model architecture:
\begin{itemize}
    \item ~\textbf{NoTime}: a variant of \textbf{LT} that neither 
 temporally shifts keyframes during data generation, nor predicts time-warps. \edit{Similar to ~\cite{oreshkin2022motion}'s formulation, \textbf{NoTime} only predicts \residual. }
    \item ~\textbf{NoWarp}: a variant of ~\textbf{LT} that uses the same dataset generation scheme, but just predicts $\Delta \mathbf{X}$, not a global time-warp.
    \item ~\textbf{IMP(C)}: imputation solution. \edit{We retrain ~\cite{tevet2023human} on 60-frame clips to be an unconditioned motion diffusion model. At inference time, we replace the prediction with input constraints at all denoising steps greater than or equal to diffusion step C. }
    \item ~\textbf{CondMDI}: \edit{Current state-of-the-art in diffusion-based motion-inbetweening~\cite{flexinbetween}, trained to exactly match the timing of the input keyframes. We retrain \textbf{CondMDI} on our 60-frame dataset and remove text conditioning during training to better match our set up. Please see the Supplement for further details. } \\
\end{itemize}

\edit{ The \textbf{CondMDI} and \textbf{NoTime} baselines evaluate the ability of an in-betweening model, trained to expect hard timing constraints, to handle keyframes with imprecise timing. The \textbf{NoWarp} and \textbf{IMP(C)} baselines explore alternative strategies for retiming these imprecisely timed keyframes: \textbf{NoWarp} utilizes our data generation approach without a dedicated retiming mechanism, while \textbf{IMP(C)} examines the effectiveness of an unconditioned diffusion prior in accurately retiming keyframes. Lastly, our method \textbf{LT} integrates a data generation strategy with an explicit warp function to handle scenarios with imprecisely timed keyframes. }

We construct test sets of approximately timed keyframes and detailed motion pairs in the same manner as our dataset generation method (Section~\ref{sec:datasetgen}): given motion clips from HumanML3D (in this case, from the test motions), slice them to \textit{F}-frame-long sequences, and choose an extrema pose at random, $\mathbf{X}_k$. We temporally shift the pose by varying $\Delta k$ to produce \blockingpose and remove frames from $[k-W, k+\Delta k)$ and $[k+\Delta k + 1, k + W)$ to produce \blocking. 

Our final test set consists of \edit{21440} $\mathbf{X}$ and $\mathbf{Y}$ pairs; our evaluation treats $\mathbf{X}$ as the model input, and the ground truth $\mathbf{Y}$ as the output that was desired when laying out keyframes in $\mathbf{X}$.

\subsubsection{Metrics}

We compare model performance on reconstruction accuracy (ability of the model to reproduce the dynamics of the entire intended motion), diversity (ability of the model the produce a variety of motions given the same input), and keypose error (ability of the model to preserve the poses of the input keyframe constraints).
 
To measure reconstruction accuracy, we measure this by computing L2 distance between the global and local joint positions (L2-Pos), velocities (L2-Vel), accelerations (L2-Acc), and jerks (L2-Jerk) of the generated vs ground truth motions. We consider a model with better reconstruction accuracies to be best aligned with our goal of generating \detailed that is similar in structure and dynamics to ground truth. \edit{We also report jitter, a common measure for overall motion quality.}

Because input constraint timings are permitted to change, we report keypose error (KPE) as the distance between the most similar pose in $\mathbf{Y}$ to \blockingpose. A full description of all metrics can be found in the Supplemental, as well as additional quantitative evaluation of our method against baselines. We show results for reconstruction accuracy, KPE, jitter, and diversity in Table~\ref{tab:reconstruction}.

\begin{table*}[t!]
\centering
    \caption{\textbf{Metrics.} We measure reconstruction accuracy between generated motions and ground truth motions. ``G'' (global) denotes a statistic calculated in world space, and ``L'' (local) is calculated local to each frame's root position. On reconstruction metrics, ~\textbf{LT} scores higher than all baselines on both low and higher-order statistics. Importantly to timing, ~\textbf{LT} most faithfully reconstructs higher-order effects, but keeps \textit{KPE} balanced. This suggests that the learned warping function is important for producing plausible and desired timing, while preserving the keyframe. An additional benefit of treating timing as a loose constraint is that ~\textbf{LT} can also achieve greater motion diversity while still keeping \textit{KPE} low. We \textbf{bold} the result with the highest performance per metric and \underline{underline} the second-highest.}
 \resizebox{\linewidth}{!}{%
 \begin{tabular}{c c c c c c c c}
  &  L2-Pos ($10^{-1}$) G/L $\downarrow$ & L2-Vel ($10^{-4}$) G/L $\downarrow$ & L2-Acc ($10^{-4}$) G/L $\downarrow$ & L2-Jerk ($10^{-3}$) G/L $\downarrow$ & KPE ($10^{-2}$) $\downarrow$ & Jitter ($10^{-2}$) $\downarrow$ & Diversity $\uparrow$ \\ 
 \midrule

  \edit{IMP(0)} & \edit{1.20 / 0.174} & \edit{57.66 / 12.36} & \edit{128.42 / 23.25} & \edit{40.81 / 70.73} & \edit{\textbf{0.00}} & \edit{1.72} & \edit{6.71} 
 \\
 \edit{IMP(1)} & \edit{1.23 / 0.183} & \edit{8.12 / 3.70} & \edit{3.08 / 1.58} & \edit{0.62 / 0.27} & \edit{1.41} & \edit{0.48} & \edit{\underline{ 6.96 }} \\
 \edit{IMP(5)} & \edit{1.35 / 0.120} & \edit{7.05 / 3.48} & \edit{2.28 / 1.29} & \edit{0.41 / 0.21} & \edit{1.68} & \edit{0.43} & \edit{\textbf{7.20}} \\
 
\edit{CondMDI} & \edit{0.43 / 0.069} & \edit{5.88 / 2.39} & \edit{5.87 / 1.52} & \edit{1.77 / 0.37} & \edit{0.120} & \edit{0.75} & \edit{2.43} \\

NoTime & \underline{0.03 / 0.018} & \underline{1.44} / 1.12 & \underline{0.71 / 0.55} & \underline{0.12 / 0.09} & \underline{0.017} & \underline{0.26} & 2.49 \\
 NoWarp & 0.04 / 0.020 & 1.63 / \underline{1.11} & 0.91 / 0.61 & 0.18 / 0.10  & 0.025 & 0.26 & 3.60 \\
 LT (Ours) & \textbf{0.03 / 0.017} & \textbf{1.35 / 1.02} & \textbf{0.60 / 0.46} & \textbf{0.096 / 0.068} & 0.019 & \textbf{0.22} & 3.68 \\
 \bottomrule
 \end{tabular} }
 \label{tab:reconstruction}
\end{table*}

\edit{Note that L2-based reconstruction accuracy metrics do not fully capture the generative nature of models. Multiple plausible predictions may exist for the same input. While exact similarity to the ground truth is not the primary objective, L2 metrics are commonly employed as a rough indicator of the overall correctness of generated motions~\cite{araujo2023circlecapturerichcontextual,athanasiou2024motionfixtextdriven3dhuman,li2023egobodyposeestimationegohead,twostagetransformers,skelbetween,goelsiggraph}. In general, it is indeed difficult to quantitatively measure quality or correctness of generative models. Therefore,  numerical results should be interpreted alongside the qualitative evaluations in our videos.  }

\subsubsection{Discussion}
\textbf{Our model produces motion with more realistic global timing.}   Because baselines cannot retime input constraints, they cannot generate motion matching the statistics of the desired motion as well as \textbf{LT}. 
Qualitatively, we notice that \textbf{CondMDI} sometimes generates motion with a global trajectory drift away from keyframe constraints (please see Supplemental for further discussion). \edit{Nevertheless, \textbf{LT} exhibits better performance in local reconstruction methods than \textbf{CondMDI} (L2-Acc 0.46 < 1.52,  L2-Jerk 0.068 < 0.37). \textbf{LT} also has the lowest amount of jitter across all baselines.}

\edit{A pure imputation-based solution, $\mathbf{IMP(0)}$, exhibits large jumps around input constraints, manifesting as high jitter and L2-Vel, Acc, and Jerk; pure imputation does not produce harmonized motions. Stopping imputation ($\mathbf{IMP(1)}$, $\mathbf{IMP(5)}$) at a late-stage diffusion step smooths out the discontinuities, but still fails to beat $\mathbf{LT}$ across all reconstruction metrics. }

While \textbf{NoTime} and \textbf{NoWarp} are not far off from \textbf{LT} on position scores, they score worse on higher order metrics in comparison (L2-Acc 0.71 > 0.6, L2-Jerk 0.12 > 0.096). This suggests that \textbf{NoTime} and \textbf{NoWarp} attempt to correct implausible timing purely with spatial detail, but in doing so, cannot generate motion that matches the dynamics and timing of the desired, ground truth motion.  This suggests that the presence of the explicit warping function is important for producing well-timed motion in our setting.

\textbf{Our model balances the trade-off between accurate timing and keypose preservation} \edit{While the pure imputation-based solution \textbf{IMP(0)} matches keyposes exactly, due to direct overwriting of predictions with input constraints until the final diffusion step, this comes at the cost of very poor timing as measured by higher-order reconstruction metrics (seen qualitatively as large discontinuities in the motion). Stopping imputation early leads to a rapid increase in KPE: the model ignores input constraints. Imputation is not sufficient for accurately retiming keyframes.}  

\edit{With no mechanism for retiming, \textbf{CondMDI} may fail to preserve an imprecisely timed keypose (KPE 0.12). While the \textbf{NoTime} baseline performs very well at preserving keyframes (KPE 0.017), this comes at the expense of timing. } \textbf{LT}, on the other hand, balances keypose preservation (KPE 0.019) while also demonstrating quantitatively higher performance on timing reconstruction statistics, i.e., acceleration and jerk. This suggests that by supplying $\textbf{LT}$ with a global time-warp, which has a lower degree of freedom than the typical pose feature representation, $\textbf{LT}$ does not have to sacrifice timing to preserve the keypose.

\textbf{Our model achieves greater motion diversity by treating timing as a loose constraint.} \edit{Motion diversity, adherence to constraints, and motion quality provide an interesting trade-off. Generative models can achieve high diversity by simply ignoring all input keyposes, or generating unrealistic/noisy motions. Ideally, however, a model should generate diverse motion while still respecting pose constraints and maintaining quality.} \edit{Similar to findings in prior work, ~\textbf{IMP(C)} methods can achieve high diversity, e.g., 7.20, but at the cost of either very high KPE (the generated motions do not adhere to the constraints), or very low reconstruction quality (the motion is highly unrealistic or disjointed)}.  In contrast, our approach balances KPE with motion diversity and quality. ~\textbf{LT}'s flexibility in timing introduces an additional degree of freedom, enabling our model to generate  diverse motions, while still adhering to keypose constraints and maintaining quality.

\section{Conclusion}
\textbf{Limitations and Future Work.} While our method can create detailed motion, the generated motion is not guaranteed to be physically accurate, e.g., limbs may slightly intersect with other body parts. A physics-based motion postprocessing approach~\cite{yuan2023physdiff} would be an interesting mechanism to incorporate. Our method does not handle finer-grained control on loose timing constraints, e.g., incorporating a combination of ``loose'' and ``hard'' timing constraints, precisely controlling how many frames a keyframe constraint is permitted to shift in time, or changing the overall length of the motion sequence. We believe that addressing these challenges is a very exciting direction for future work and would provide even more flexible control over character motion. 

\section{Acknowledgments}
Purvi Goel is supported by a Stanford Interdisciplinary Graduate Fellowship. We thank the anonymous reviewers for constructive feedback; Vishnu Sarukkai, Sarah Jobalia, Sofia Di Toro Wyetzner, and Mia Tang for proofreading; James Hong, Zander Majercik, and David Durst for helpful discussions; Meta for gift support.

\bibliographystyle{eg-alpha-doi} 
\bibliography{motion_unblocking}   

\end{document}